\documentclass[11pt,preprint]{aastex}
\usepackage{natbib}
\usepackage{amsmath,amssymb}
\usepackage{mathptmx}
\usepackage{graphicx}
\usepackage[flushleft]{threeparttable}
\usepackage{longtable}
\usepackage[font=small,labelfont=bf]{caption}
\usepackage{graphicx,rotating,booktabs}
\usepackage{makecell}
\usepackage[verbose]{placeins}
\usepackage{blindtext}
\usepackage{graphicx,rotating,booktabs}
\usepackage{amsmath}
\newcommand  \ergs     {\ifmmode {\rm erg\,s}^{-1} \else erg s$^{-1}$\fi}
\newcommand  \Msunyr     {\ifmmode {\rm \Msun\,yr}^{-1} \else \Msun\ yr$^{-1}$\fi}
\newcommand  \msunyr     {\ifmmode {\rm \Msun\,yr}^{-1} \else \Msun\ yr$^{-1}$\fi}
\newcommand  \cmii     {\ifmmode {\rm cm}^{-2} \else cm$^{-2}$\fi}
\newcommand  \cmiii     {\ifmmode {\rm cm}^{-3} \else cm$^{-3}$\fi}
\def\Hubble{\ifmmode {\rm km\,s}^{-1}\,{\rm Mpc}^{-1}\else km\,s$^{-1}$\,Mpc$^{-1}$\fi}
\def\Msun{\ifmmode M_{\odot} \else $M_{\odot}$\fi}
\def\msun{\ifmmode M_{\odot} \else $M_{\odot}$\fi}
\def\Lsun{\ifmmode L_{\odot} \else $L_{\odot}$\fi}
\def\Zsun{\ifmmode Z_{\odot} \else $Z_{\odot}$\fi}
\def\qo{\ifmmode q_{0} \else $q_{0}$\fi}
\def\Ho{\ifmmode H_{0} \else $H_{0}$\fi}
\def\ho{\ifmmode h_{0} \else $h_{0}$\fi}
\def\qo{\ifmmode q_{0} \else $q_{0}$\fi}
\def\ao{\ifmmode a_{0} \else $a_{0}$\fi}
\def\to{\ifmmode t_{0} \else $t_{0}$\fi}
\def\omm{\ifmmode \Omega_{{\rm M}} \else $\Omega_{{\rm M}}$\fi}
\def\omlam{\ifmmode \Omega_{\Lambda} \else $\Omega_{\Lambda}$\fi}

\def\mgii{\ifmmode {\rm Mg}{\textsc{ii}} \else Mg\,{\sc ii}\fi}
\newcommand \MgII {\ifmmode {\rm Mg}\,{\sc ii}\,\lambda2798 \else Mg\,{\sc ii}\,$\lambda2798$\fi}
\def\Hbeta{\ifmmode {\rm H}\beta \else H$\beta$\fi}

\def \Lop{$L_{5100}$}
\def \L3000a{$L_{3000}$}
\def \L1450{$L_{1450}$}   
\def \Cf {$C_f$}
\newcommand{\lbol}  {\ifmmode L_{\rm bol} \else $L_{\rm bol}$\fi}
\newcommand{\Lbol}  {\ifmmode L_{\rm bol} \else $L_{\rm bol}$\fi}
\newcommand{\lagn}  {\ifmmode L_{\rm AGN} \else $L_{\rm AGN}$\fi}
\newcommand{\LAGN}  {\ifmmode L_{\rm AGN} \else $L_{\rm AGN}$\fi}
\newcommand{\LUV}   {\ifmmode L_{\rm 1350} \else $L_{\rm 1350}$\fi}
\newcommand{\Ldust}   {\ifmmode L_{5\mu m} \else $L_{5 \mu m}$\fi}
\newcommand{\lsf}   {\ifmmode L_{\rm SF} \else $L_{\rm SF}$\fi}
\newcommand{\LSF}   {\ifmmode L_{\rm SF} \else $L_{\rm SF}$\fi}
\newcommand{\LTOR}   {\ifmmode L_{\rm torus} \else $L_{\rm torus}$\fi}
\newcommand{\LIR}   {\ifmmode L_{\rm IR} \else $L_{\rm IR}$\fi}
\newcommand{\LHD}   {\ifmmode L_{\rm HD} \else $L_{\rm HD}$\fi}
\newcommand{\lledd} {\ifmmode L/L_{\rm Edd} \else $L/L_{\rm Edd}$\fi}
\newcommand{\Ledd} {\ifmmode L/L_{\rm Edd} \else $L/L_{\rm Edd}$\fi}
\newcommand{\fwmg}  {\ifmmode {\rm FWHM}\left(\mgii\right) \else FWHM(\mgii)\fi}
\newcommand{\CFHD}  {\ifmmode {\rm CF}_{\rm HD} \else ${\rm CF}_{\rm HD}$\fi}
\newcommand{\mbh}   {\ifmmode M_{\rm BH} \else $M_{\rm BH}$\fi}
\newcommand{\MBH}   {\ifmmode M_{\rm BH} \else $M_{\rm BH}$\fi}
\newcommand{\mstar}   {\ifmmode M_{*} \else $M_{*}$\fi}
\newcommand{\Mstar}   {\ifmmode M_{*} \else $M_{*}$\fi}

\def  \mic         {$\mu$m}
\def  \MgII         {\ifmmode {\rm Mg}\,{\sc ii}\,\lambda2798
                  \else Mg\,{\sc ii}\,$\lambda2798$\fi}
\def  \mgii         {\ifmmode {\rm Mg}\,{\sc ii} \else Mg\,{\sc ii}\fi}
\def\ha{\ifmmode {\rm H}\alpha \else H$\alpha$\fi}
\def\Ha{\ifmmode {\rm H}\alpha \else H$\alpha$\fi}
\def\La{\ifmmode {\rm L}\alpha \else L$\alpha$\fi}

\def \spitzer      {{\it Spitzer}}
\def \wise {{\it WISE}}
\def \WISE {{\it WISE}}

\def \herschel {{\it Herschel}}
\def \Herschel {{\it Herschel}}
\def \chandra {{\it Chandra}}

\def\Chisq{\ifmmode \chi^{2} \else $\chi^{2}$}
\def \zzz {\ifmmode z=2-3.5 \else $z =2-3.5$\fi} 

\def\zzzz {$z \simeq$4.8}
\def\z48{$z \simeq$4.8}
\def\z33{$z \simeq$3.3}
\def\z24{$z \simeq$2.4}

\begin{document}

\title{Star formation black hole growth and dusty tori in the most luminous AGNs at \zzz}

\author{Hagai Netzer\altaffilmark{1},
Caterina Lani\altaffilmark{1},
Raanan Nordon\altaffilmark{1},
Benny Trakhtenbrot\altaffilmark{2},
Paulina Lira\altaffilmark{4}
\& Ohad Shemmer\altaffilmark{3}
}

\altaffiltext{1}
{School of Physics and Astronomy and the Wise Observatory,
The Raymond and Beverly Sackler Faculty of Exact Sciences,
Tel-Aviv University, Tel-Aviv 69978, Israel}

\altaffiltext{2}
{Institute for Astronomy, Department of Physics, ETH Zurich, Wolfgang-Pauli-Strasse 27, CH-8093 Zurich, Switzerland
(Zwicky postdoctoral fellow)}
\altaffiltext{3}
{Department of Physics, University of North Texas, Denton, TX 76203, USA}

\altaffiltext{4}
{Departamento de Astronomia, Universidad de Chile, Camino del Observatorio 1515, Santiago, Chile}

\email{netzer@wise.tau.ac.il}


\begin{abstract}
We report  \herschel/SPIRE observations of 100 very luminous, optically selected active 
galactic nuclei (AGNs) at \zzz\ with $\log$~\LUV\ (\ergs)$\ge 46.5$,
where \LUV\ is $\lambda L_{\lambda}$ at 1350\AA. The distribution in \LUV\ is similar to
the general distribution of SDSS AGNs in this redshift and luminosity interval.   
We measured star formation (SF) luminosity, \LSF, and SF rate (SFR) in 34 detected sources by fitting combined SF and 
torus templates, where
the torus emission is based on \wise\ observations. We also obtained statistically significant stacks for the undetected sources in two luminosity groups.
The sample properties are compared with those of very luminous AGNs at $z>4.5 $.
The main findings are:
1) The mean and the median SFRs of the detected sources are $1176^{+476}_{-339}$ and $1010^{+706}_{-503}$ \msunyr,
respectively. The mean SFR of the undetected sources is $148$ \msunyr.
The ratio of SFR to BH accretion rate is $\approx 80$ for the detected sources and less than 10 for the undetected sources.
Unlike a sample of sources at $z\simeq 4.8$ we studied recently, there is no difference in \LAGN\ and only a very small difference in \LTOR\ between detected and undetected sources.
2) The redshift distribution of \LSF\ and \LAGN\ for the most luminous, redshift 2--7 AGNs are different. Similar to previous studies,
the highest \LAGN\
are found at $z\approx 3$. However, \LSF\ of such sources peaks at $z \approx 5$. Assuming the objects in our sample are hosted
by the most massive galaxies at those redshifts, we find that approximately 2/3 of the hosts are already below the main-sequence 
of SF galaxies at z=2-3.5.
3) The spectral energy distributions (SEDs) of dusty tori at high redshift are similar to the shapes found  in low redshift, low luminosity
AGNs. \herschel\ upper limits put strong constraints on the long wavelength shape of the SED ruling out several earlier suggested
torus templates are applicable for this sample.
4) We find no evidence for a luminosity dependence
of the torus covering factor in sources with 
$\log$~\LAGN\ (\ergs)=44-47.5. This conclusion is based on the recognition that the estimated \LAGN\ in several earlier studies is highly uncertain and
non-uniformally treated. 
The median covering factors over this range are $0.68 $ for isotropic dust emission and $0.4 $ for anisotropic emission.

\end{abstract}

\section{Introduction}
\label{sec:introduction}

The comparison of black hole (BH) growth and star formation (SF) across cosmic time 
is essential for the understanding of the parallel evolution of galaxies and super-massive BHs. This topic has received much attention
over the last decade, in particular since the launch of ESA Herschel Space Observatory \cite[][hereafter \herschel]{Pilbratt2010}.
This mission provided superb far infrared (FIR) capability and hence deeper and more systematic study of SF rate
(SFR), specific SFR (sSFR), and SF luminosity (\LSF) at high redshift. 
It is now possible to compare \LSF\ 
and AGN bolometric luminosity (\LAGN) over a large redshift range in thousands of
sources at $z<1$ and hundreds of sources at $z>2$. At $z\sim 1$, \herschel/PACS, with bands at 70, 100 and 160\mic,
 provides the most reliable far-IR (FIR) fluxes, and thus \LSF, for
the host galaxies of AGNs with 
\LAGN$\geq 10^{45}$\ergs. For $z>2$, \herschel/SPIRE \citep{Griffin2010}, with bands at 250, 350 and 500\mic, is more efficient and detection is limited  
only by confusion noise which translates to $\log$~\LSF\ (\ergs)$\sim 45.6$ \ergs\ ($\sim 100$ \msunyr) at $z=2-3$. 

Studies of AGN hosts show that at all redshifts most AGNs reside in SF galaxies 
\cite[e.g.][]{Silverman2008,Mainieri2011,Santini2012,Mullaney2012,Rosario2012}. Moreover, 
there is no evidence for different host properties between systems with active or dormant BHs that have the same stellar mass (\Mstar) \citep{Rosario2013}.
There is already an extensive literature on the comparison of \LSF\ and \LAGN\ at all redshifts with somewhat 
ambiguous conclusions \cite[e.g.][and references therein]{Netzer2009a,Shao2010,Hatziminaoglou2010,Rosario2012,Harrison2012,Page2012,
Mullaney2012,Chen2013,Hickox2014,Stanley2015}. Some of the differences depend on the selection method,
 FIR (which is basically SF selection) or X-ray (preferentially AGN selection). It also depends on the treatment of 
undetected objects (stacking, statistical analysis of upper limits, etc) that are usually the majority of the sources at high redshift.
A general result that emerged from these studies is
that  the \LSF-\LAGN\ plane can be divided
into two regimes with very different distributions. The first is the ``SF dominated'' regime where \LSF$>$\LAGN. Here 
the correlation between \LSF\ and \LAGN\
depends critically on the selection and averaging methods. In studies like \cite{Rosario2012} and \cite{Stanley2015}, selection and binning is by X-ray flux and stacking by
FIR flux. In this case, there is no apparent correlation between \LSF\ and \LAGN. 
In contrast, FIR selected samples \citep{Mullaney2012,Chen2013,Delvecchio2015},
where binning is in FIR flux and stacking by X-ray flux, show a clear correlation between log(\LSF) and log(\LAGN) with a slope very close to 
unity.  
In the second ``AGN dominated'' regime, where \LAGN$>$\LSF, the sources seem to cluster around a power-law line of the form \LSF$\propto$\LAGN$^{0.7}$.
 There are a few theoretical attempts to explain these correlations based on the different
duration of SFR and BH accretion \citep{Neistein2014}, the different duty cycles \citep{Hickox2014}, and the nature of SF and BH accretion in 
merging galaxies \citep{Thacker2014,Volonteri2015a,Volonteri2015b}.

This work is a continuation of \cite{Netzer2014} which discusses \LAGN\ and \LSF\ in a flux limited sample of optically selected  $z \simeq 4.8$ 
AGNs. These objects are the most luminous AGNs that are powered by the most massive BHs at this redshift.
Out of the 44 AGNs in that sample, 10 were detected by \herschel/SPIRE and stacking of 29 undetected sources gave statistically significant
fluxes at all  SPIRE bands. Five more sources gave ambiguous results. \cite{Netzer2014} compared \LSF, \LAGN\ and BH mass (\MBH) in the detected and undetected
source and showed that the detected sources, with
the higher \LSF, are also those with the higher \LAGN\ and the more massive BHs. The \herschel-detected sources are located close to the border line between the
two regimes mentioned above with \LAGN$\sim$\LSF. The related work of \cite{Leipski2014} provided similar data for a large 
number of luminous, randomly selected AGNs  at $z>5$. 
In their study, the detection  limit is very high, due to the high redshift.
As a result the number of directly detected sources is less than 10 and there is no statistically meaningful way to  compare \LSF, \LAGN\ and BH mass.

The present work follows our earlier study of high redshift, luminous AGNs. It
presents data on  the most luminous type-I AGNs at $z=2-3.5$ observed by \herschel\ using deep SPIRE observations.
This means that we avoid several studies of very high luminosity type-II AGNs like the \cite{Castinani2015} and \cite{Drouart2014} works
on radio loud sources.
We focus on two central themes of AGN research. The first is SF in the host galaxies of these
sources and the second near-Infrared (NIR) and mid-infrared (MIR) dust emission in the vicinity of the central BHs.
The latter is related to AGN tori; a topic which was studied, extensively, since the mid-1980s (see \cite{Antonucci1993} for review of
the earlier results and ideas, and the recent review by \cite{Netzer2015} where newer developments and more recent references are provided). 
One way to study such tori
is to compare the observed NIR-MIR luminosity, \LTOR, with \LAGN\ and derive the ``torus covering factor'' as a function of source luminosity. This covering factor 
is directly related to the fraction of type-I and type-II AGNs at different redshifts 
\cite[e.g.][and references therein]{Treister2008,Lusso2013,Roseboom2013,Merloni2014}. Here we extend these studies to the highest possible luminosity
and focus on the
spectral energy distribution (SED) emitted by the torus, the anisotropy of the torus radiation, and the way to determine the covering factor.

The structure of the paper is as follows:
In \S2\ we describe the sample and the Wide Field Infrared Survey Explorer (\WISE) and \herschel\ data. \S3 provides a detailed
explanation of the method used to separate SF and torus emission in \herschel-detected and undetected sources.
We use our new data, and those from the literature, to discuss SF and AGN emission at $z>2$ and to provide new information
about torus emission and covering factor at all luminosities and redshifts.  \S4 presents the conclusions of our work.
Throughout this paper we assume $\Ho=70$ \Hubble, $\omm = 0.3$ and $\omlam = 0.7$.
For the conversion to SFR we assume that \LSF\ is obtained from integration over the 8--1000\mic\ range, and 1\msunyr\ corresponds to 
a slightly rounded value of $10^{10}$ \Lsun\
based on a Chabrier initial mass function \citep{Chabrier2003} as adopted by \cite{Nordon2012}.

\section{Sample Selection Observations and Data Analysis}
\label{sec:sample}

\subsection{Sample selection}
\label{sec:sample_selection}
The sample selected for this study consists of type-I AGNs found in the data-release 7 (DR7) of the Sloan Digital Sky 
Survey \cite[SDSS][]{Abazajian2009} and observed
by \Herschel/SPIRE \cite[see][]{Griffin2010}.
The chosen redshift range
is 2--3.5 and the chosen luminosity $\log \lambda L_{\lambda}$(1350 \AA\ \ergs)$\geq 46.5 $ (hereafter \LUV).
Sixteen objects of the entire sample have been observed by \Herschel\ as part of a dedicated open-time cycle 2 project (PI: H. Netzer). 
Five of these sources are not SDSS AGN.
The remainder are obtained from the \Herschel\ archive.
This includes targeted and serendipitous sources, as well as
objects in several large GTO surveys.  
We used the \cite{Shen2011} catalog that lists all bright AGNs with SDSS spectra, and searched the \Herschel\ archive for sources that exceed the chosen \LUV\
in that catalog (3383 sources) and were observed by \herschel.  
The search includes AGNs 
observed with SPIRE in a ``Small Scan'' mode 
and sources that are located in relatively large fields observed by various key projects. 
We searched for such sources in the following extra-galactic fields: H-ATLAS \citep{Eales2010}, COSMOS \citep{Scoville2007}, 
Extended Groth Strip (EGS, \citealt{Davis2007}), 
ELAIS \citep{RR1999}, Lockman Hole \citep{Lonsdale2003}, Strip-82 \citep{AM2007} and  B\"{o}otes \citep{Jannuzi2004}.
This list does not cover
all fields observed with SPIRE and reflect the publicly available \herschel\ data correct to late 2014.
For these fields we use the catalogs and publicly released images and the standard pipeline.
The total number of sources identified in this way is about 120. 
Only a handful 
of the selected sources were observed by \Herschel/PACS and all but one were not detected. Since the small number is not sufficient for stacking analysis,
we decided not to use the PACS observations. 
since their inclusion for only a small fraction of the sample, will affect the uniformity of the sample in an undesirable way.
We only consider observations where the SPIRE confusion limits
have been reached. 
With SPIRE resolution, the 1-$\sigma$ flux limits correspond roughly to 5.8 mJy at 250\mic, 6.3 mJy at 350\mic, and  6.8 mJy at 500\mic.
Some candidates were located behind lensing clusters. We decided not to include lensed sources due to the complications associated with the lensing.

Of the originally selected sources, we decided to remove about 20. One source was removed because of a neighboring source less
than 5 arc-sec away
All other sources that were removed from the original list are not detected by \herschel\ and are situated very close to the edge of their filed,
or in other areas with large gaps in the \herschel\ scans. There are clear indications that the
confusion limits have not been reached in these cases.
 The total number
of \herschel-observed sources remained after this procedure is 100.
Out of these, 34 are considered detections with 
$3 \sigma$ flux detection in 3 (31 sources) or 2 (250\mic\ and 350\mic, 3 sources) of the SPIRE bands. The rest have 
upper limits in all three bands (see \S~\ref{sec:flux_measurements}).
Given the relatively large range in torus luminosity which affects the measurements of 
\LSF\ (\S~\ref{sec:stacking})
we divided the sources into two sub-groups based on their \LUV. The first group consists of objects with
$ 46.5 \leq \log$~\LUV\ (\ergs)$\leq 46.7$ and the second $46.7< \log$~\LUV\ (\ergs).  There are 57 (19 detected and 38 upper limits) AGNs in the former
group and 
43 (15 detected and 28 upper limits) in the latter group.

\begin{figure}[t]
\centering
 \includegraphics[height=8.5cm]{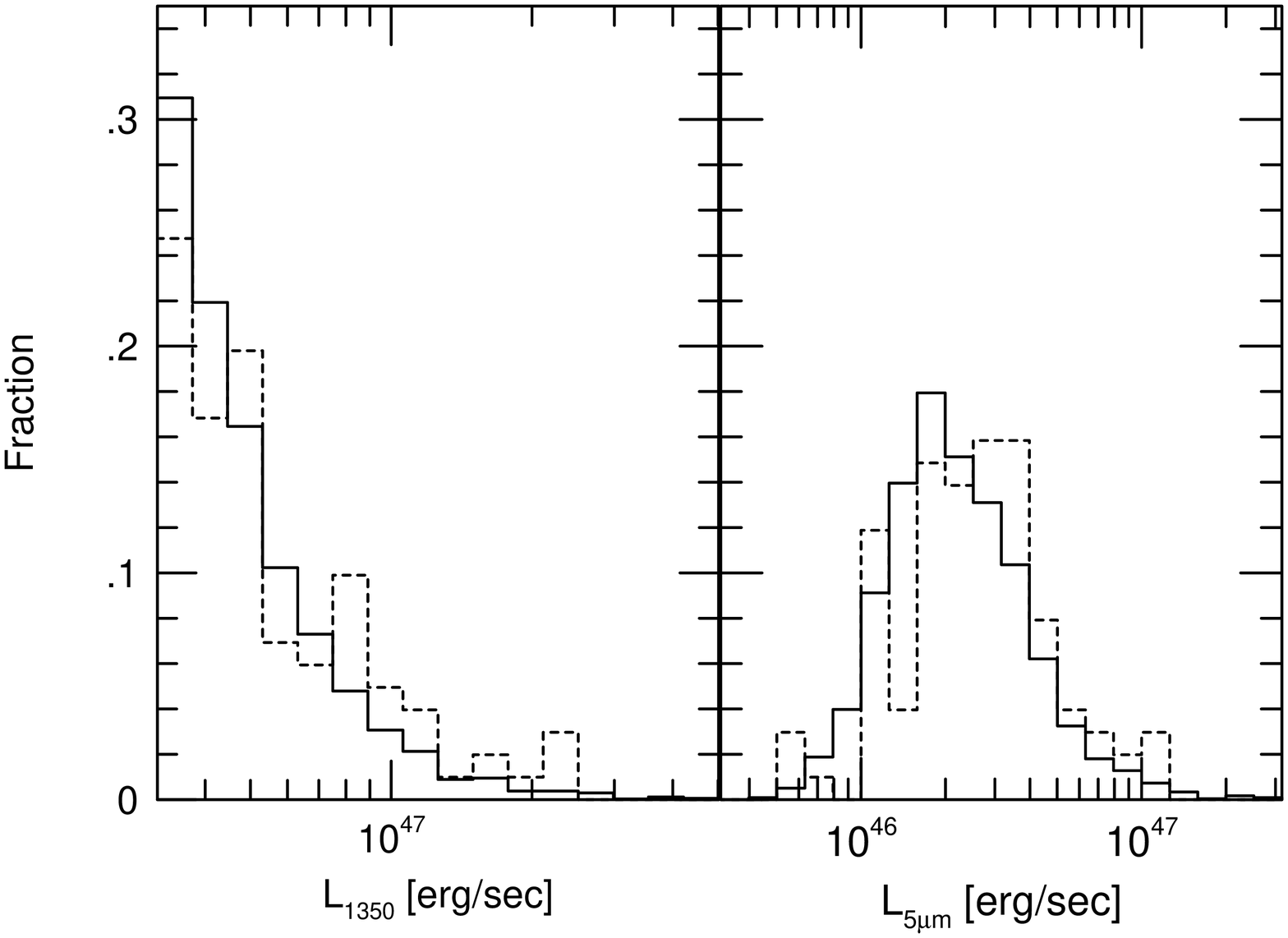}
 \caption{
Left: A comparison of the bolometric luminosity distribution of the \herschel-observed AGNs with $\log$~\Lbol\ (\ergs)$\geq 46.5$ 
and \zzz\ (dashed line)
 with the distribution of all SDSS AGNs from \cite{Shen2011} in the same luminosity and redshift range (solid line).
Right: The same for \Ldust.
}
\label{fig:sdss_distributions}
\end{figure}

The \Herschel\ observations are part of several different observing programs, with different goals, and are not covering the SDSS part of the sky in a systematic
way. We therefore checked how representative of the general population these sources are. To do this we compared the luminosity distribution of our 100
with all type-I AGNs from the \cite{Shen2011} catalog covering the same 
\LUV\ and redshift range. The comparison is shown in Fig.~\ref{fig:sdss_distributions}. 
We tested the null hypothesis that the two distribution are drawn from the same parent population using the  
standard, two distributions KS test. The resulted probability, $p=0.071$, suggests that the null hypothesis cannot be ruled out.
We also compared the redshift distributions of the two samples across the chosen range of 2--3.5. The distributions are
very similar with almost identical median and a KS probability of $p=0.93$. 
 Separating into luminosity groups we find a median redshifts of 2.57 for the low luminosity group and 2.50 for the high luminosity group.

\subsection{\herschel\ flux measurements}
\label{sec:flux_measurements}

The fluxes in the small scan SPIRE maps were measured using the {\it sourceExtractorSussextractor} task in the \Herschel\ software package HIPE
\citep[for details of the algorithm and parameters see][]{Savage2007}.
This task extracts fluxes for all sources detected in the image simultaneously, and thus also partially solve for confusion/blending with nearby sources.
In the first iteration, the task was run as a blind extraction on the 250~\mic\ image, with the detection threshold set to 3~$\sigma$
plus the target's optical coordinates as priors.
If no source was found within 5\arcsec\ from the optical prior location, a second iteration was run using the source list from the first iteration.
The flux extraction from
 the 350 and 500\mic\ images was done using the coordinates of the blindly detected 250\mic\ sources as priors.
For all our sources, there were no cases of 350\mic\ blind detections without 250\mic\ counterparts.
The 500\mic\ images are much more affected by confusion and a list of priors is always necessary for their extraction.
In the large fields of major SPIRE projects 
(H-ATLAS SDP described in \citealt{Pascale2011} and \citealt{Rigby2011} , HerS described in \citealt{Viero2014}, HerMES \citealt{Oliver2012})
we take the fluxes provided in the source catalogs as released by the 
corresponding teams,  using a 3-5'' search radius around the SDSS coordinates.

\subsection{Stacking of non-detected Herschel sources}
\label{sec:stacking}

The majority of our \Herschel-observed sources (68\%) were not detected in any of the  SPIRE bands.
In such cases we used stacking analysis which was done separately for 
the two luminosity subgroups.
 The stacking procedure is identical to that used by \citet{Nordon2012} and has been thoroughly tested in previous works.
It is based on many previous works in the literature, such as \citet{Bethermin2010} and \citet{Dole2006}.
In the present work we obtained both mean and median stacks for the two luminosity groups. 

The depth of the SPIRE fields is usually limited by confusion due to multiple/bright nearby sources, rather than by instrument noise 
The SPIRE confusion limits are approximately 5.8, 6.3 and 6.8 mJy at 250, 350 and 500\mic\, respectively \citep{Nguyen2010}.
Stacking the images on the (optically measured) position of otherwise undetected sources, reduces the confusion roughly by (0.5--1)$\sqrt N$ 
and increases the accuracy of the mean (or median) flux accordingly.
 For stacking we use {\it residual images} that were created by removing all sources detected by sourceExtractorSussextractor (using the standard tools within the HIPE package).
This procedure does not remove any flux from the (undetected) stacked sources and serves only to somewhat flatten the background.
Working on residual images makes it easier to 
determine the background level and lowers some of the confusion-noise in the final stacked image.
Residual images were constructed for all fields used in this work.
The images must use the same pixel scale, therefore we re-projected those images with different scales onto a new grid.
The background level is determined by creating a histogram of pixel fluxes and locating its peak.
The removal of the bright sources from the residual image already contributed to removing the high values tail in the histogram, making 
the distribution more symmetric around its peak.
The peak of the distribution represents the most likely background flux value of a randomly chosen pixel. This flux is subtracted from the image before stacking
(i.e. the new pixel-flux histogram is centered on zero).

For each stacking position we project the residual image onto a small grid (``stamp'') with the optical source coordinates centered on the central pixel.
The stamps are collected into a cube, where each stamp is multiplied by the redshift correction factor (see \S~\ref{sec:k_correction}).
We then take the mean (or median) value of a cube column 
(same pixel in all stamps) as the value for the corresponding pixel in the stacked image.
Thus, the final stacked images represent the mean or median fluxes for our sample.
 We also create background stacks, where we stack on random locations in the residual images, selected to be at least several PSFs away from the 
location of the optical source.
The number of background stamps is four times the number of stacked sources, but from this pool of stamps each bootstrap step (see \S~\ref{sec:Bootstrap}) 
draws a number of stamps similar to that of the real sources stacks.
By duplicating in parallel every step done with the real sources stack, the background stacks serve as control and an estimator of the noise level (instrumental and confusion),
as well as of biases due to inaccuracies in background subtraction.

The flux in a final mean or median stacked image is measured from the peak of a fitted 2D gaussian to the source in the center of the image with a sub-pixel allowance for 
adjustments in peak location and FWHM. These minor adjustments 
are necessary because the stamps are taken from different \Herschel\ images, which may potentially have slightly different systematic offsets in 
coordinates relative to the optical positions.
 The flux in the background stacked image is measured in an identical way and is used as an estimator to the systematic error due to non-zero background (verified to be smaller than the random errors by factor of a few)
and the significance of detection (see \S~\ref{sec:Bootstrap}).
 Median stacks that were used in this work are shown in Fig. A1. 

\subsubsection{Redshift and K-corrections}
\label{sec:k_correction}
Our stacking analysis requires mixing fluxes from different redshifts, hence we need to assign an effective redshift to the stacked source.
If we were comparing a sample of detected sources, we would apply a redshift k-correction to each source to account for the different rest frame wavelengths of the SPIRE bands,
and a correction due to the different luminosity distances.
For the stacks we apply the same procedure (all redshifts are known), but we do so to  
each stamp in the cube before taking the means or the medians. This was done assuming a stacked redshift of $z=2.5$, 
which is very close to the median redshifts listed above.
For the redshift range chosen (2--3.5), the k-correction and the luminosity distance correction tend to cancel each other out, and
the combined factor is very close to 1 (0.85--1.35 with a median of 1.1 for the 250\mic\ stacks and an even narrower range for the longer wavelength stacks)

For the k-correction we use a warm ULIRG template ($\log(L_{\rm IR}/L_{\odot}) > 12.7$) from the \cite{Chary2001} library (the exact template shape makes only a little difference). 
This translates to:
\begin{equation}
 k = \frac{ F_{\nu}(\lambda_{\rm filter}/(1+2.5)) }{ F_{\nu}(\lambda_{\rm filter}/(1+z)) } \, ,
 \label{eq:k_correction}
\end{equation}
where $F_{\nu}$ is the template flux as a function of rest frame frequency and $\lambda_{\rm filter}$ is the observed wavelength of the band.
We note that the slope of this template at wavelengths longer than the FIR peak is quite close to the slope of the \cite{Mor2012}  torus SEDs (a topic which
is discussed in detail in \S~\ref{sec:torus_properties}).
Thus, for most combinations of SPIRE filters and source redshifts in our sample, there is little difference whether we assume a (relatively) warm star-formation
dust emission, or a pure torus emission.
In any combination of wavelength and redshift (worst being at $z=3.5$), the difference between using a torus template and a ULIRG template for the k-correction is less than 10\%,
and in most cases much lower around the median redshift of 2.5.
As for the luminosity distance, we corrected the fluxes to $D_{\rm L}(z=2.5)$.

\subsection{Bootstrap uncertainties}
\label{sec:Bootstrap}
We use a standard bootstrap method to estimate the uncertainty on the stacked fluxes of our sources.
From our cube of $N$ stamps, we randomly re-sample $N$ stamps allowing repetition.
In parallel we select $N$ stamps from the background stamp cube.
The new cubes are then treated in the exact same way as the originals and a mean or median flux is measured.
The process is repeated 500 times to  obtain a distribution of re-sampled fluxes.
The central 68\% of the distribution was taken as the $\pm 1 \sigma$ limits on the measured medians.
The limits represent a combination of noise and the intrinsic spread of source fluxes within the sample.
The corresponding lower and upper limits obtained from the background stack are used to estimate the significance of detection of the stacked source -
i.e., ruling out the null hypothesis that the stacked source is a fluctuation (a combination of instrumental and confusion noise).

\subsection{\WISE\ measurements and Torus SED}
\label{sec:torus_sed}

All our sources were detected by \WISE\ \citep{Wright2010} in at lease one of its four bands: 3.4(W1), 4.6(W2), 12(W3), and 22(W4)\mic. 
General discussions of the most luminous \wise-detected AGNs are given in \cite{Weedman2012} and \cite{Vardanyan2014} that include all the sources in our sample.
 The above references also detail the standard data extraction procedure that we used in the present work.
For the chosen redshift range, the four bands cover the rest-frame 1--7\mic\ wavelength range and are employed to determine
the torus SED and covering factor. 
 
The torus SED template we use throughout this paper is the one suggested by \cite{Mor2012}. This SED is based on fitting \spitzer\ spectra,
after subtracting a SF contribution,  and NIR photometry of approximately  100 
 nearby, type-I AGNs. 
The data were fitted with a three component model: 
a hot ($\sim 1400-1800$K) dust source, assumed to represent the inner walls of the torus where only graphite grains
can survive the local flux of the central source, a warm graphite+silicate torus based on the calculations of \cite{Nenkova2008b}, and dust emission from the NLR.
This  template was extended to the FIR by combining
it with a 100K, $\beta=1.5$ grey body at $\lambda > 35$\mic. 
The SED is roughly flat, in $\lambda L_{\lambda}$, between 2 and 25\mic\ and then drops sharply at longer wavelengths.
\cite{Mor2012} discussed the uncertainty on the shape of the SED and its dependence on source luminosity (see their Fig. 6).
The study of \wise/SDSS AGNs by \cite{Vardanyan2014} shows that the very flat SED between 2 and 10\mic\ is a common properties of the most luminous AGNs at high redshift. 

An earlier paper by \cite{Mullaney2011} proposed
 somewhat different,  luminosity dependent SEDs, turning-down at longer wavelengths in the range of 30-40\mic. 
The work is based on fitting broken power law models to MIR-FIR data after subtracting the SF contribution. 
The main differences between the two works are the different assumptions
about the SF contribution to the observed MIR emission, which in turn can influence the derived SED shape at long wavelength,
the assumed clumpy torus vs. a broken power law, and the treatment of the very hot dust. 
As shown in \S~\ref{sec:torus_sed},  some of the \cite{Mullaney2011}
SEDs seem to be in conflict
with our \herschel\ observations. A different torus model proposed by \cite{Polletta2006} and \cite{Polletta2007}, was used by \cite{Tsai2015} to study extremely luminous
high redshift AGNs and compare them with the most luminous galaxies discovered by \wise. The 2--20\mic\ part of this SED is in good agreement with the \cite{Mor2012} template
but its long wavelength part is very different and is also in conflict with our \herschel\ observations discussed in \S~\ref{sec:torus_sed}.
A large part of the difference must be due to the neglect of SF contribution to the long wavelength part of this template that
was assumed to be AGN-dominated.

Several recent works 
\citep{Roseboom2013,Lusso2013,Assef2013,Lira2013,Leipski2014} adopted different fitting methods. \cite{Roseboom2013} used
 various combinations of a hot blackbody,
and a warm dusty torus to fit broad-band NIR-MIR data in a large sample of \WISE-selected sources.
\cite{Lusso2013} used broad band JHK and \spitzer\ MIR  data to fit 513 XMM-COSMOS sources. Their assumed torus SED is more empirical, based
on observations of low luminosity AGNs. The approach adopted by \cite{Leipski2014} is similar
to the \cite{Roseboom2013} method. Finally, \cite{Lira2013} fitted local type-II AGNs with several different theoretical predictions. 

The NIR emission of dusty tori in type-I AGN can vary considerably from one source to the next. The variation is stronger than that
observed in the MIR part, with some 10-20\% of the sources showing much weaker
2-4\mic\ dust emission. Such objects, here referred to as ``weak-NIR'' AGNs, have been investigated in detail in several papers 
\cite[e.g.][]{Mor2011,Mor2012,Roseboom2013,Leipski2014}.
The SED adopted here is inappropriate for the weak-NIR sources.

 The prescription from our preferred torus model (see \S3.4), corresponds to  \LTOR=$(3.58^{+0.69}_{-0.4})$\Ldust, where   
\Ldust\ stands for  $\lambda L_{\lambda}$ at 5\mic, \LTOR\ is the observed integrated SED over the 1--200~\mic\ range, and the upper and lower
limits are the values corresponding to the 25 and 75
percentiles in \cite{Mor2012}. 
As explained in \S~\ref{sec:torus_properties}, this is a very good assumption for all the objects
in our sample except for the weak-NIR sources ( 12 objects). Obviously  the total energy radiated by the torus can differ substantially 
from \LTOR, due to anisotropic dust emission (\S~\ref{sec:covering_factor}).

To complete the comparison with the general population at $2 \leq z \leq 3.5$, we used the entire SDSS sample from the \cite{Shen2011} catalog, with the
above mentioned redshift and luminosity cuts, and obtained \wise\ data for all these sources. We calculated the luminosities in the W3 (12\mic) and
W4 (22\mic) windows and used them to estimate \Ldust. We only consider sources with 3$\sigma$ detections in the W1, W2 and W3 bands (3217 objects
 out of the total of 3383).
For sources with both W3 and W4 3$\sigma$ detections (72\% of the cases), we used a linear interpolation in 
$\log(L)$  to 
estimate \Ldust. For the remaining sources, with upper limits on the flux in W4, we assume that \Ldust\ equals the W3 luminosity.
Given the very similar luminosities in the W3 and W4 bands (which we verified for all sources with detections in both bands), this is a very good approximation
of \Ldust\  given our choice of the torus SED. 
The results of the comparison
are shown as two histograms in Fig.~\ref{fig:sdss_distributions}. The histograms look similar and the probability of the 
two-distributions KS test is $p=0.035$.
More information about the luminosity distribution of SDSS AGNs observed by \wise\ is given in \cite{Vardanyan2014}.

Table~1 provide basic information about the sample and Table~ 2 lists  the newly obtained fluxes and luminosities.
The median and the mean stack fluxes are given in Table~3.

\subsection{\LAGN\ estimates: bolometric correction factors}
\label{sec:bolometric}

Two of the central issues discussed in this paper, the correlation between \LAGN\ and \LSF, and that  between \LAGN\ and 
\LTOR, depend
on the method used to estimate \LAGN. This requires the use
of bolometric correction factors applied to the observed continuum luminosity at different wavelengths. The bolometric correction factor has been a 
point of some confusion in earlier
studies and, hence, requires more explanation \cite[e.g.][]{Marconi2004,Richards2006,Runnoe2012a,Runnoe2012b,Trakhtenbrot2012,Krawczyk2013}.
Since \LUV\ is directly probing the AGN accretion power, we will use this wavelength to estimate \LAGN.

The work of \cite{Marconi2004} suggests that all bolometric correction factors, at all wavelengths, decrease with increasing \LAGN. 
This general trend has been
confirmed in later works but the actual factors are rather different. 
For example, at the very high luminosity end ($\log$~\LUV\ (\ergs)$\sim 47$), the bolometric correction factor suggested by \cite{Shen2011} for \LUV\ is 3.8, the one
by \cite{Trakhtenbrot2012} $\sim 2$, that by \cite{Runnoe2012a} $\sim 3.23$, and the one by \cite{Krawczyk2013} less than 2 (extrapolating from their
calculations at 2500\AA).  
All numbers quoted  here refer to isotropic emission at all wavelengths.
As shown in \S~\ref{sec:discussion}, these differences can affect, considerably, some of the conclusions about AGN tori.
\cite{Runnoe2012b} is a systematic study of the bolometric correction factors in the range 1.5--24~\mic.
The numbers at 3 and 7\mic\ are basically identical and can be translated to the wavelength of interest here (5\mic) as a bolometric correction
factor of approximately $ 8.5 \pm 0.8$.
This gives \Ldust/\LUV$\simeq 0.38$ for $\log$~\LUV\ (\ergs)=47. 
In our sample the median value is \Ldust/\LUV$\simeq 0.44$ with a 25--75 percentile range of 0.36--0.55. 
Thus, at the high luminosity end, our numbers and those of \cite{Runnoe2012b} are in very good agreement.

The approach we adopt here is to use the same bolometric correction factor for all our high luminosity sources. 
Our bolometric correction factors are taken from \cite{Trakhtenbrot2012} and are  based on a large sample of
SDSS AGNs and the inter-calibration of continuum luminosities at three wavelength bands, 5100\AA, 3000\AA\ and 1400\AA.
The scaling is a combination of the luminosity-dependent correction factors suggested by \cite{Marconi2004} 
which were checked and verified over a large range of luminosity and redshift.
Of particular importance in high redshift objects is $Bol_{1350}$, the bolometric correction factor applied to \LUV.
\cite{Trakhtenbrot2012} show that for high luminosity AGNs $Bol_{1350} \sim 2$ with a rather large scatter,  the order a factor 2
(but note that only 230 sources from the large sample were available
for the analysis). This estimate is independent of \LUV, therefore we also 
experimented with a second approximation: $Bol_{1350}=49-$\LUV) which give bolometric correction factors between 2.5 and 1.6 in 
the present sample and is also consistent with the results of \cite{Trakhtenbrot2012}.  bolometric correction factor 
is in somewhat better agreement with thin accretion disk models of 
\cite[e.g.][]{Capellupo2015}.
As shown in \S~\ref{sec:covering_factor}, such differences can lead to different conclusions
concerning the estimates of the torus covering factor.

\subsection{Combined torus and star-formation SEDs}
\label{sec:composite_sed}

At the redshifts considered here, the three SPIRE bands cover roughly the rest frame 50--170~\mic. This wavelength range is 
the region where the torus emission drops
rapidly towards longer wavelength and where the SF emission should peak.
The FIR luminosity of many of our sources is considerably below their NIR-MIR luminosity but the torus contribution at FIR energies must be taken
into account properly in order not to be confused with the SF emission. 
We take this into account by fitting  both the \WISE\ and SPIRE photometry using a composite SED that includes a torus and SF components.
We first fit the torus template to the \WISE\ data using the rest-frame fluxes between 1.5-6\mic.
We do not take into account systematic changes in the shape of the torus partly because there is no spectroscopic information about the
spectral shape at high luminosity and partly
because our own analysis appears to justify the chosen SED at $\lambda>25$~\mic\ (see \S~\ref{sec:torus_properties}).
We fit the template to the data using a simple sum-of-squares minimization.
The logarithmic uncertainty on the scaling of the template is determined from:
\begin{equation}
 \sigma_{\rm scale}^2 =  \frac{t^2}{N(N-1)} \sum_{\rm filter} \left( \log{\nu}L_{\nu}({\rm filter}) - \log{\nu}L_{\nu}({\rm template}) \right)^2 \, ,
\end{equation}
where $t$ is the student-t test correction factor required to correct the standard deviation estimate to represent 68\% when the number of points (N) is small.
In almost all cases, the uncertainties on the torus luminosities are very small.
Note that these are formal errors based on the assumption of an identical torus SED for all sources.  Even small changes in the SED, like those described in Mor and Netzer (2012),  will result in additional scatter which can be significantly larger than the scaling uncertainties used here.

\begin{figure}[t]
\centering
\includegraphics[width=6.5cm]{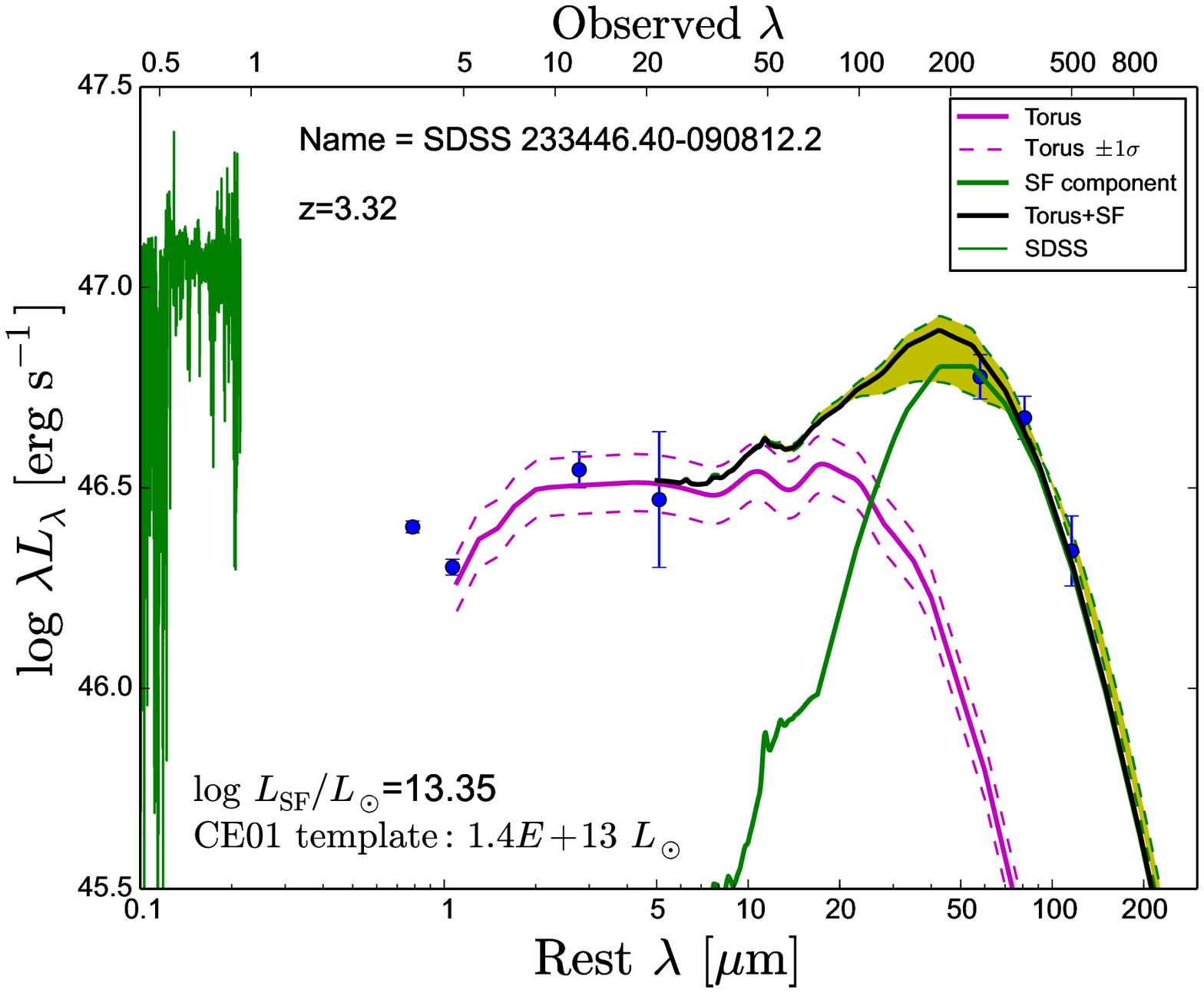}
\includegraphics[width=6.5cm]{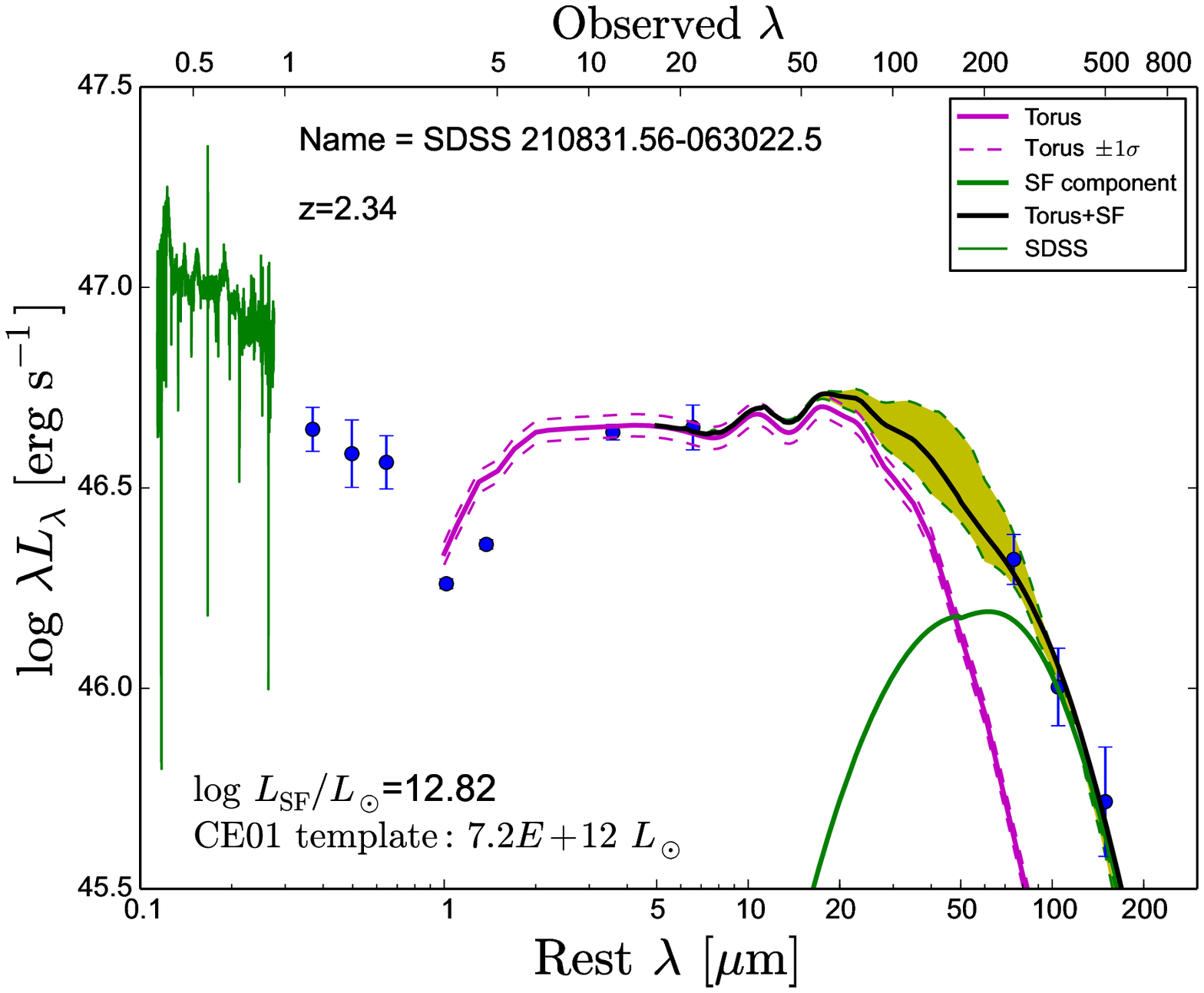}
  \caption{Examples of combined SED fittings for objects with different \Ldust/\LSF. 
SDSS data are shown in green and NIR (J,H,K) and MIR (\wise) data as blue points. The shaded area shows the range of SF templates used in the fit.
The key to the various curves used in the fit is given in the top right
of each panel.
  }
  \label{fig:fit_1}
\end{figure}

Once the SED scaling is determined, we subtract the torus contribution from the flux in the SPIRE bands and add (in quadrature) the torus scaling uncertainty 
to the SPIRE photometric errors.
The torus-subtracted fluxes are then fitted with a SF galaxy template from the \cite{Chary2001}  SED library.
The fit 
allows some flexibility in both dust temperature (template shape) and scaling (total IR luminosity) by considering all the templates
in the library whose 250/(1+z)~\mic\ luminosity is in the range of $\pm 0.5$ dex from the observed value. In all cases, PAH contributions to the W4 \wise\ band
are extremely small, of order 1\%, and hence do not affect the pre-determined torus SED.
We then minimize $\chi^2$ in the SPIRE bands and select the 
template with the lowest value. We prefer this method to the one involving all 5 bands since the $\chi^2$ values in the W3 and W4 bands are
significantly smaller than those in the SPIRE bands which will severely bias the combined analysis.
We search for the combination of \cite{Chary2001} template and scaling that produces the lowest/highest \LSF\ within $\chi^2 < \chi^2_{\rm min}+1$
and take these values as 
our low/high \LSF\ 1-$\sigma$ limits listed in Table 2..
Fig.~\ref{fig:fit_1} shows two examples of combined fits for \herschel-detected sources with different
SF templates. 

\begin{figure}[t]
\centering
\includegraphics[height=6cm]{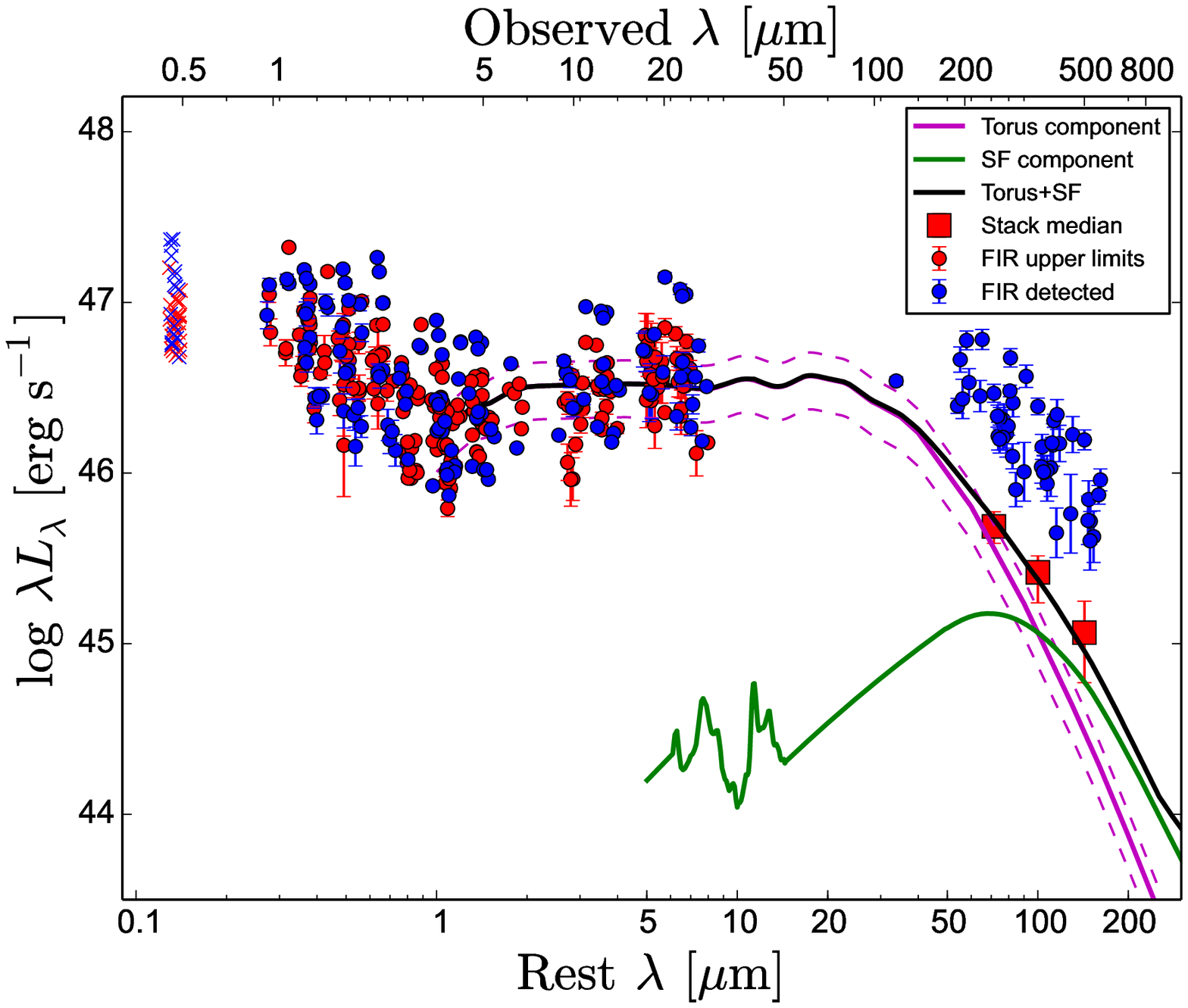}
\includegraphics[height=6cm]{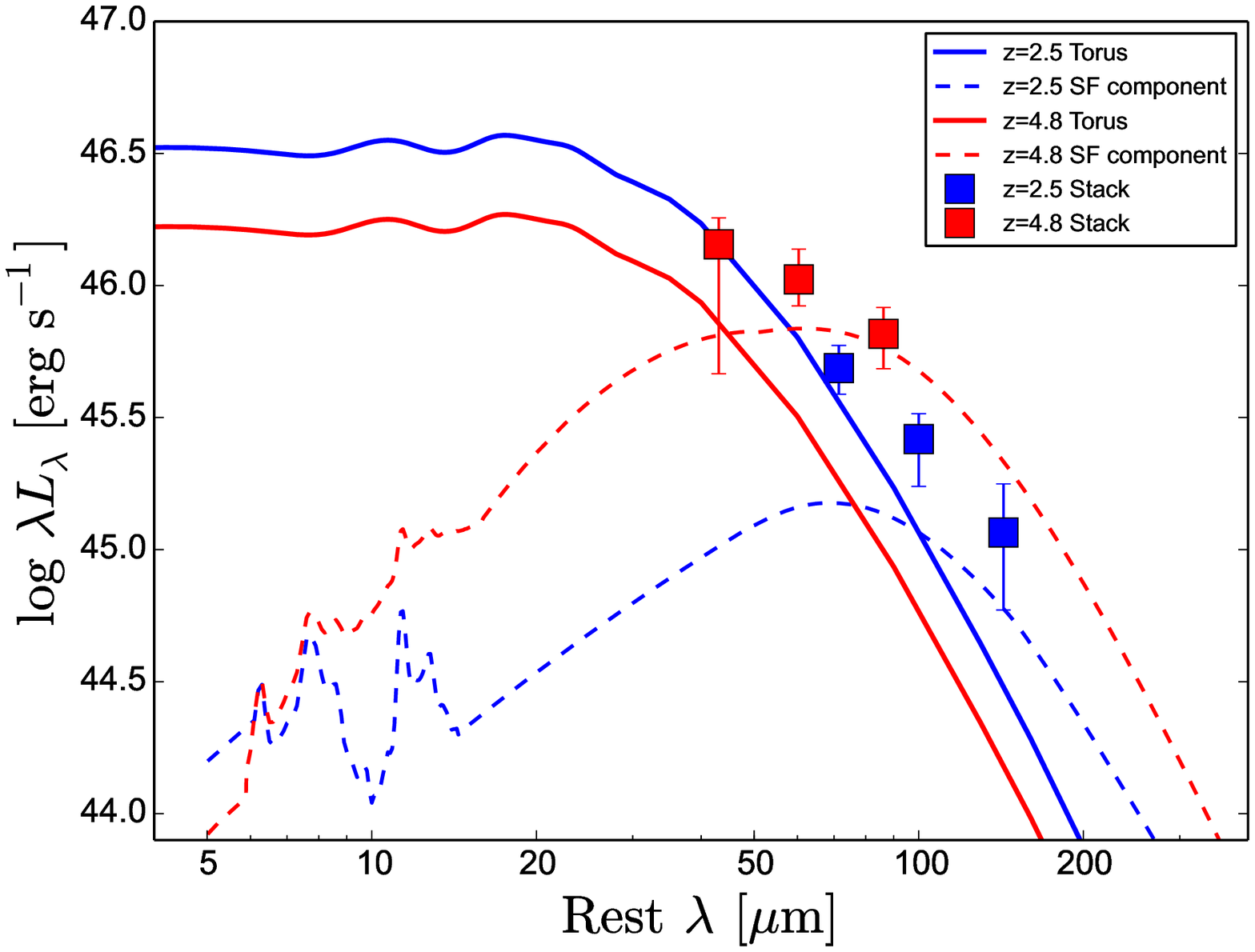}
  \caption{
Left: Summary of the observed properties of the sources in the high luminosity group ($\log$\LUV$> 46.7$).. 
``X'' symbols denote the observed \LUV. Magenta line is the
median \cite{Mor2012} torus and green line the SF template fitted to the median stack. 
\herschel-detected sources are shown in blue and undetected in red (circles for upper limits and squares for the median stack).
All symbols are also shown in the upper insert.
Note the great similarity between detected and undetected sources over the entire wavelength range, except for the FIR.
Right: A comparison of the torus and FIR stacks in the present sample (blue) and the $z \simeq 4.8$ \cite{Netzer2014}
sample (red). Note that the SFR in the higher redshift sample is larger while the mean torus luminosity is smaller.
  }
  \label{fig:stack_L1}
 \end{figure}

We followed a similar procedure to fit the stacked spectra. Since we have \wise\ data for all the sources, we 
can create two median torus templates for the two subgroups as defined by the threshold luminosity \LUV..
The scatter in torus luminosity in the two groups is small and this procedure results in well defined torus templates.
The two are processed separately since their median \Ldust\ differ by approximately 0.3 dex which has an effect on the combined
torus-SF fit. 
We measured the stack \LSF\ in a manner similar to the one used to fit the detected sources.
The stack SPIRE fluxes are only 2--4 times larger than the predicted torus fluxes and the uncertainties on the resulted SFRs are large. 
As in the case of the detected sources, we tried both a ``free-shape'' approach (all the SEDs in the relevant range of the template libraries) and the more conventional
way of using the actual luminosities in the SF library. 
Fig.~\ref{fig:stack_L1} shows the fitted SED of the median stacked spectrum of the more luminous \zzz\ sub-sample together with the \WISE\ data. 
The formally measured SFR 
is  74 \msunyr\ with an acceptable range 47--100 \msunyr. 
The application of the second approach (i.e. luminosities obtained directly from the template library) 
results in SFR within 20\% of this number. 
Finally we repeated the same procedure to obtain mean stacks for the two luminosity groups.

Fig.~\ref{fig:stack_L1}  also shows the same method applied to the \cite{Netzer2014} $z\simeq 4.8$ sample. The fits quality is similar but there is a significant
difference in the ratio \LSF/\LTOR. While the higher redshift sample is of lower AGN luminosity, and hence lower \LTOR, its mean SFR is higher by a factor of 
approximately 3. The lower redshift of the present sample allow us to go significantly
below the SFR of the most massive SF galaxies at \zzz\ \cite[e.g.][]{Schreiber2015} while for the \zzzz\ sample, the mean SFR
of the undetected sources is very similar to that expected for the most massive SF hosts at this redshift.
We come back to these points in \S~\ref{sec:discussion}

\section{Results and Discussion}
\label{sec:discussion}

The new observations presented here show several characteristics of the most luminous AGNs in the universe, in particular the relationships
 between SFR in the host, the intrinsic AGN luminosity, and dust emission by the torus.
In this section we provide a detailed discussion of the new results and 
compare them  with those obtained for other samples of high-luminosity AGNs.

 The left panel of Fig.~\ref{fig:stack_L1}
provides a visual summary of many of the properties of the \herschel-detected and undetected sources in the sub-sample with $\log$~\LUV\ (\ergs)$>46.7$.
The diagram demonstrates the similarity of \herschel-detected and
undetected sources across the UV-optical-NIR-MIR range, and the large range in \LSF\ for these sources. 
The median and mean SFRs for the various groups, and the entire sample, are summarized in
Table~4\footnote{Note that we do not provide proper uncertainties on the mean and median SFRs of the entire sample
since we did not stack all the undetected sources together. The numbers in the last column are simple means of the other columns.}

\subsection{Torus and star formation emission at $z>5$}
A recent detailed discussion of SF and AGN emission at very high redshift is given in \cite{Leipski2014}. This work presents data for  69
 $z>5$ high luminosity
AGNs observed by \spitzer, \herschel, and various sub-mm telescopes. The sample was not selected in a systematic way and it is not clear how well
it represents the $z>5$ AGN population. The total number of FIR-detected sources is 7-11, depending on the number
of \herschel\ bands required to define a source as ``FIR-detected'' (Leipski et al. require detection in 4 \herschel\ bands, from both PACS and SPIRE).
\cite{Leipski2014} used a three-component dust emission model to measure what they defined as hot-dust NIR blackbody emission, warm
dust torus emission, and cold dust  SF emission. Since their way to model the IR SED is rather different from ours, we used
the data in their paper, and our assumed torus SED, to re-measure the FIR and dust emission in all their \herschel/SPIRE-detected sources.  
Two objects in
their samples have only 250\mic\ detections, two others both 250 and 350\mic\ detections, and 7 with detections in all three bands. We have followed our 
fitting procedure, this time with PACS and \spitzer\ data given in \cite{Leipski2014},  
and the same torus model, to derive \LSF\ and \LTOR\ for these 11 sources. We could not find a satisfactory solution for the 250\mic-only sources, and for one of the sources 
with both 250 and 350\mic\ detections. For the other 8 sources we found reliable \LSF\ estimates that, for the 7 sources in common, are similar to the \LSF\
found by \cite{Leipski2014}. Five of the eight sources belong in the category of weak-NIR AGN, a much larger fraction than in the general population.
Given the very small number of \herschel-detected sources, 
we consider this to be only tentative evidence for a larger fraction of weak-NIR sources at high redshift. 

Regarding the estimate of \LAGN, we used \Lop\ as  listed in \cite{Leipski2014} with a bolometric correction factor of 4 which is consistent with our
bolometric correction factor for \LUV\ \citep{Trakhtenbrot2012}. 
This method gives values that are 30--40\% higher than the those estimated by them from 
direct integration over the rest-frame wavelength 0.1--1\mic. The two estimates are consistent since the \cite{Leipski2014}
estimates do not take into account the luminosity of the $\lambda <0.1$\mic\ part of the SED which is included in our estimates of \LAGN.

\subsection{\LAGN\ and \LSF\ at \zzz}

We have looked for possible correlations between \LAGN\ and \LSF\ in all the AGNs discussed in this paper, and compared 
them with various correlations suggested in the literature. 
In doing so we take into consideration the fact that the methods
used for selecting the samples (e.g. X-ray selection vs. FIR selection) greatly influence the results. This was
explained in detail in \S1 where we also gave many relevant references. Since our sample is optically selected, the ones more relevant to the present study are those
based on the AGN properties, e.g. X-ray flux.

First we focus on correlations suggested  for 
``AGN dominated'' systems, i.e. those with \LAGN$>$\LSF. Such correlations were  discussed in \cite{Netzer2009a} and \cite{Netzer2014}
and can be expressed as: 
$\log L_{SF}=0.7 \log L_{AGN}+12.9$. This relationship is an attempt to fit, by eye, 
a combination of a large number of low luminosity AGNs and a small number of 
highly luminous AGNs. It is not based on a formal
regression analysis since the number of low luminosity sources far exceeds the number of very luminous systems thus biasing any attempt
to obtain a meaningful correlation that spans 4--5 orders of magnitude in luminosity. 
It is also obtained for individual sources and does not involve mean or median properties.
The second comparison is with the \cite{Stanley2015} sample of X-ray detected AGNs at redshifts 0.2--2.5. In this case, most of the objects
are not detected by \herschel\ and the authors used a survival analysis to obtain mean FIR fluxes. This approach is similar to the one used
by \cite{Rosario2012}, who used stacking of \herschel\ data, and the results of the two studies are in good agreement except, perhaps, at the
highest \LAGN\ end.  The data we compared with our observations are those found in the
redshift range $1.87< z < 2.08 $ which correspond to the most luminous objects in these samples.
Finally, we also examine 
the correlation for ``SF dominated'' sources suggested by
\cite{Delvecchio2015} for their $0.15<z<2.3$ sources. This correlation can be expressed as: 
$\log L_{AGN}=-9.32+1.18 L_{SF}$. 
It is based
on a systematic study of FIR, \herschel-selected sources in the COSMOS and GOODS fields where SF luminosities were obtained from
measured FIR SEDs and \LAGN\ from stacked X-ray (\chandra) data ($\sim$10\% detections and $\sim 90$\% undetected sources). 
The results of this work are in good agreement with the
earlier results of \cite{Chen2013} for SFR$>1$~\msunyr.

In order to use consistent estimates of \LAGN, we re-calibrated the values obtained by \cite{Netzer2014} for their $z\simeq 4.8$ sample
by adopting the approximation used here of \LAGN=2\LUV. The new values are  within 20\% of the numbers 
 found by \cite{Netzer2014}. 
The X-ray based estimates of \LAGN\ for other samples considered here 
were obtained from the original papers
using the bolometric correction factors from \cite{Marconi2004}.
We prefer this direct approach rather than the one used by \cite{Stanley2015} who first
 compared the X-ray and MIR luminosities and then used correlations between
MIR luminosity and \LAGN.

\begin{figure}
\centering
\includegraphics[height=10cm]{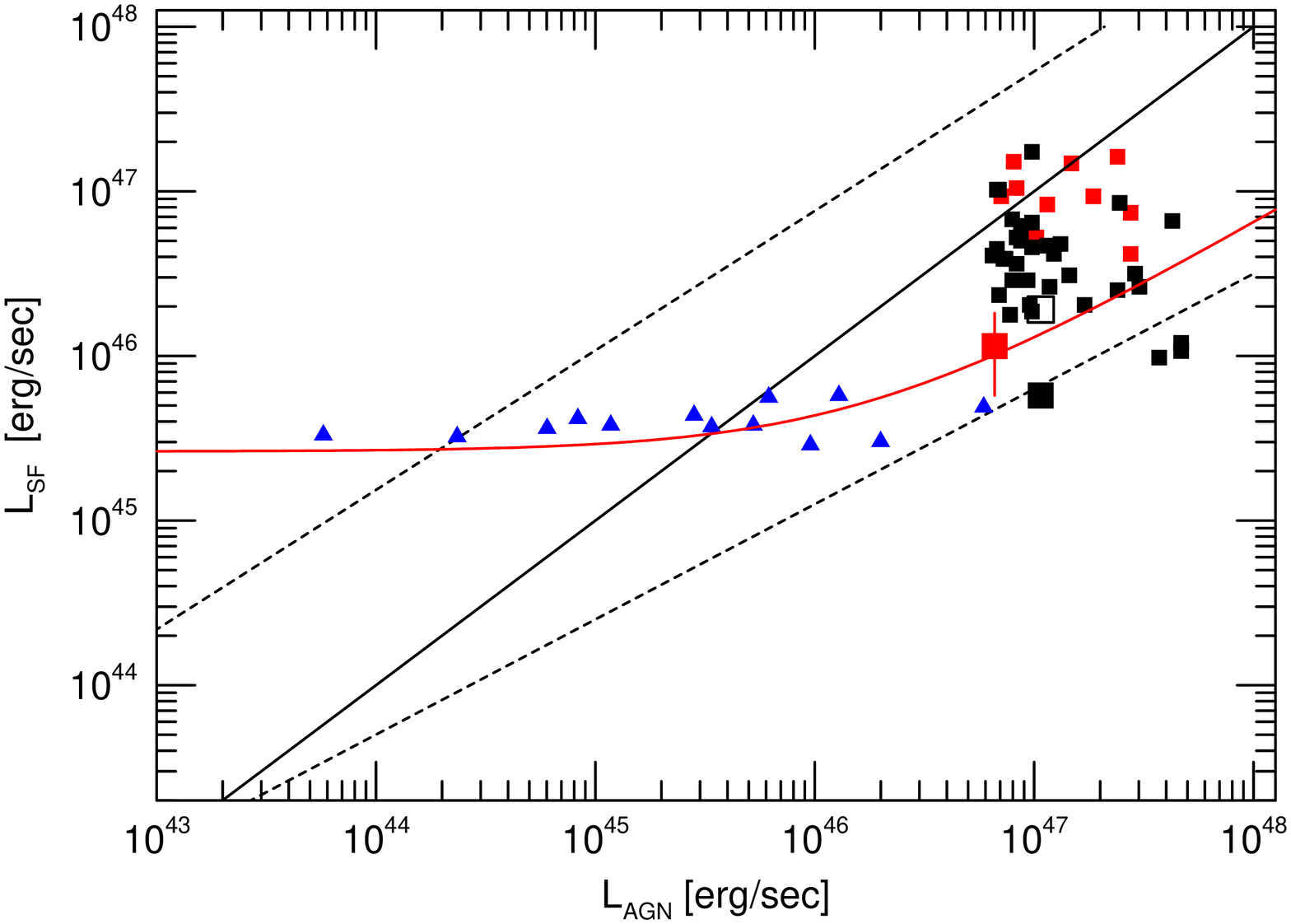}
  \caption{\LSF\ vs. \LAGN\ for various representative samples at low and intermediate redshifts, and our two 
high redshift samples at $z=2-3.5$ and $z\simeq 4.8$.
Blue triangles: \cite{Stanley2015}.
Full black squares: the present sample (the big square is the mean of all the undetected sources).
Open black square: the mean of the entire sample.
Red squares: \cite{Netzer2014} (the big square represent stacked source). 
Red solid line: the \cite{Rosario2012} fit to the  $1.5<z<2.5$ AGNs scaled up by a factor 2.
Upper dashed black line: the \cite{Delvecchio2015} fit to their SF-dominated sources.
Lower dashed line: \cite{Netzer2009a} fit to AGN-dominated sources.
Solid black line: \LSF=\LAGN.
}
  \label{fig:LAGN_LSF}
 \end{figure}

 Fig.~\ref{fig:LAGN_LSF} shows \LSF\ vs. \LAGN\ for various representative samples 
and our own samples for $z=2-3.5$ and $z\simeq 4.8$. The additional samples 
are from \cite{Delvecchio2015}, 
 where we only show the best fit line,
the \cite{Stanley2015} sample, the curve fitted by \cite{Rosario2012} to their $1.5<z<2.5$ group of sources scaled up by a factor 2  
(\cite{Rosario2012} used the measured 60~\mic\ luminosity and not \LSF), and the fit to the \cite{Netzer2009a} relationship .
For the present sample, we show all individually detected sources, the mean for the stacked sources, and the mean for the entire sample.

The above comparison is illuminating. It shows that most of the $z>2$ AGN hosts (those in the stacks) have roughly the same
\LSF\ as the mean measured in the lower redshift samples that include detected sources. 
However, \LSF\ of the \herschel-detected sources exceed this
 value by about an 
order of magnitude. As a result, the mean \LSF\ in our sample (shown as an open black square) 
is significantly higher than the typical values obtained by \cite{Stanley2015} over a similar redshift range. 
The main reason for this is probably related to the fact that very high \LAGN\ 
objects are missing from studied based on small fields,  like COSMOS, that do not sample properly the high end of the AGN
luminosity function.

\subsection{\LAGN\ and \LSF\ across time}

The high redshift samples studied here can be compared with earlier works addressing stellar mass growth via SF at
similar epochs. In particular, we want to compare the newly measured SFRs with studies of the main sequence
 (MS)
of star-forming galaxies and the IR luminosity function (IRLF) of high redshift galaxies.
Out of the numerous papers published on the correlation of stellar mass, SFR and sSFR,  
\cite[e.g.][and references therein]{Daddi2007,Wuyts2011a,Rodighiero2011,Speagle2014}
we chose to compare our results with those of \cite{Schreiber2015} that cover a very large range of stellar masses for all redshifts between
0 and 5. This study is based on FIR measurements and thus avoids the uncertainties associated with UV dust attenuation. It is also
quite complete in terms of high mass galaxies and is in good agreement with the systematic work of \cite{Speagle2014} that include 
a detailed comparison and 
calibration of different methods. For the definition of the MS, we use the parameterization given in Eq.~9 of 
\cite{Schreiber2015} that suggests a continuous growth of sSFR with redshift for the most massive galaxies, up to $z \sim 4$.
The conversion factor from SFR to \LSF\ used by these authors  is a factor of 1.7 larger than the one used here, due to the different
assumed IMF, and the following calculations take this into account.
 Finally, we assume that the width of the MS, close to its high-mass end, is $\pm 0.3$ dex, a value which is consistent with most of the 
references listed above but is, at such high redshifts, quite uncertain.

There are fewer papers about LFs
at high redshift \cite[see][and references therein]{Madau2014}. The specific work used here is \cite{Gruppioni2013} that focus on
the infrared LF (IRLF) and is based on \herschel\ observations. The highest redshifts considered in this paper are $z \sim 4$. Stellar mass functions
for these fields are shown by \cite{Schreiber2015} (Fig.~3 in their paper) and are basically identical to the ones presented 
by \cite{Ilbert2013} and \cite{Madau2014}

As explained, the $z=2-3.5$ and $z\simeq 4.8$ samples represent well the population of the most luminous 
AGNs in these  redshift intervals. 
The $z>5$ sample was chosen in a different way, and is less complete in this respect.
However, it includes many of the most luminous objects 
in this redshift range and hence provides some indication on the overall distribution in \LAGN\ and \LSF.
The overall range in \LAGN\ of the three high redshift samples is about a factor of 10. 
The range in \LSF\ is more difficult to define since approximately 70\% of the sources are not detected by
\herschel. 
Using our individual detections, and the mean and median stacks, we find this range to
be approximately  1.6 dex. Obviously the stacks may include completely quenched hosts.

\begin{figure}
\centering
\includegraphics[height=10cm]{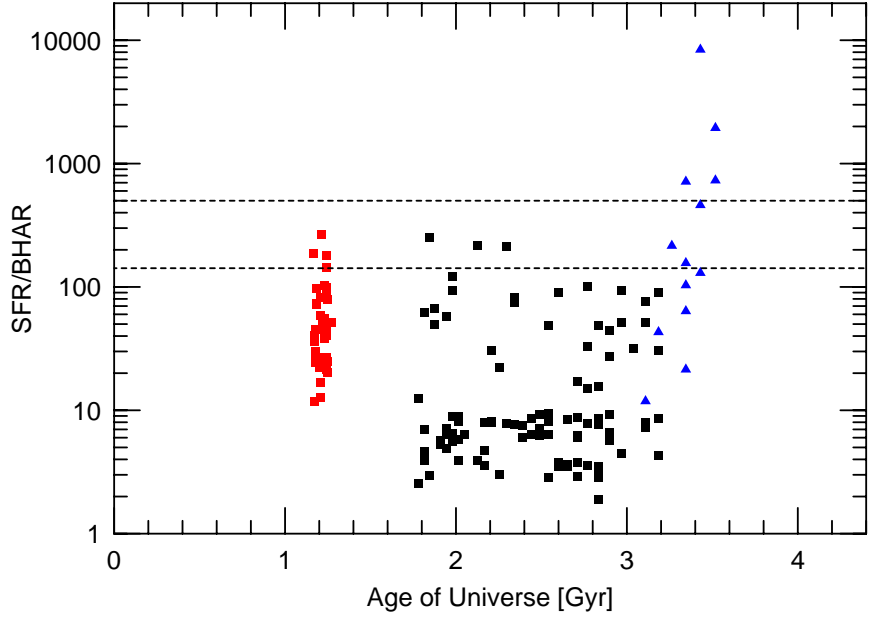}
  \caption{SFR/BHAR as a function of cosmic time for the sources and samples shown in Fig.~\ref{fig:LAGN_LSF}, using the same symbols
and colors. In this diagram we show {\it all} individual sources from the stacks assuming the corresponding mean SFRs.
 The dashed horizontal lines mark the two ratios explained in the text, SFR/BHAR=500 (``typical'' of local AGNs)  and SFR/BHAR=142
(\LSF$\simeq$\LAGN).
}
  \label{fig:SFR_BHAR}
 \end{figure}

We first consider the ratio of SFR to BH accretion rate (BHAR)
as a function of cosmic time for the SPIRE detected sources in the redshift ranges 2--3.5, $\approx 4.8$ and $>5$.
To do this we assume a mass-to-radiation conversion efficiency of $\eta=0.1$ corresponding to thin accretion disks with a spin parameter of $\sim 0.7$,
consistent with the high spin values for very massive BHs measured by \cite{Capellupo2015}.
The results are shown in Fig.~\ref{fig:SFR_BHAR} where the individual points have the same meanings as in Fig.~\ref{fig:LAGN_LSF}.
We mark in the diagram the line corresponding to SFR/BHAR=500, which is the ratio of stellar-mass to BH mass in the local universe for
galaxies hosting BHs with \mbh$\sim 10^8$~\msun\ with bulge mass which is half the total mass of the galaxy
\cite[][and references therein]{Kormendy2013}.
All sources in all three high redshift samples, except for some of the stacked averages in the \cite{Stanley2015} sample, are well below
this line. The detected sources cluster around SFR/BHAR$\simeq 80$, 
not far from SFR/BHAR=142 which corresponds to \LSF$\simeq$\LAGN\ (also marked in the diagram). 
This accretion rate ratio is very similar to the stellar-to-BH mass ratio for the most massive spheroidal galaxies in the local universe.
The undetected sources are all 
below SFR/BHAR=10.
We also note that the study of type-II radio-loud AGNs by \cite{Drouart2014}, which is based on very different ways of estimating
BHARs, shows values of SFR/BHAR that are consistent with the ones shown here.

\begin{figure}
\centering
\includegraphics[height=10cm]{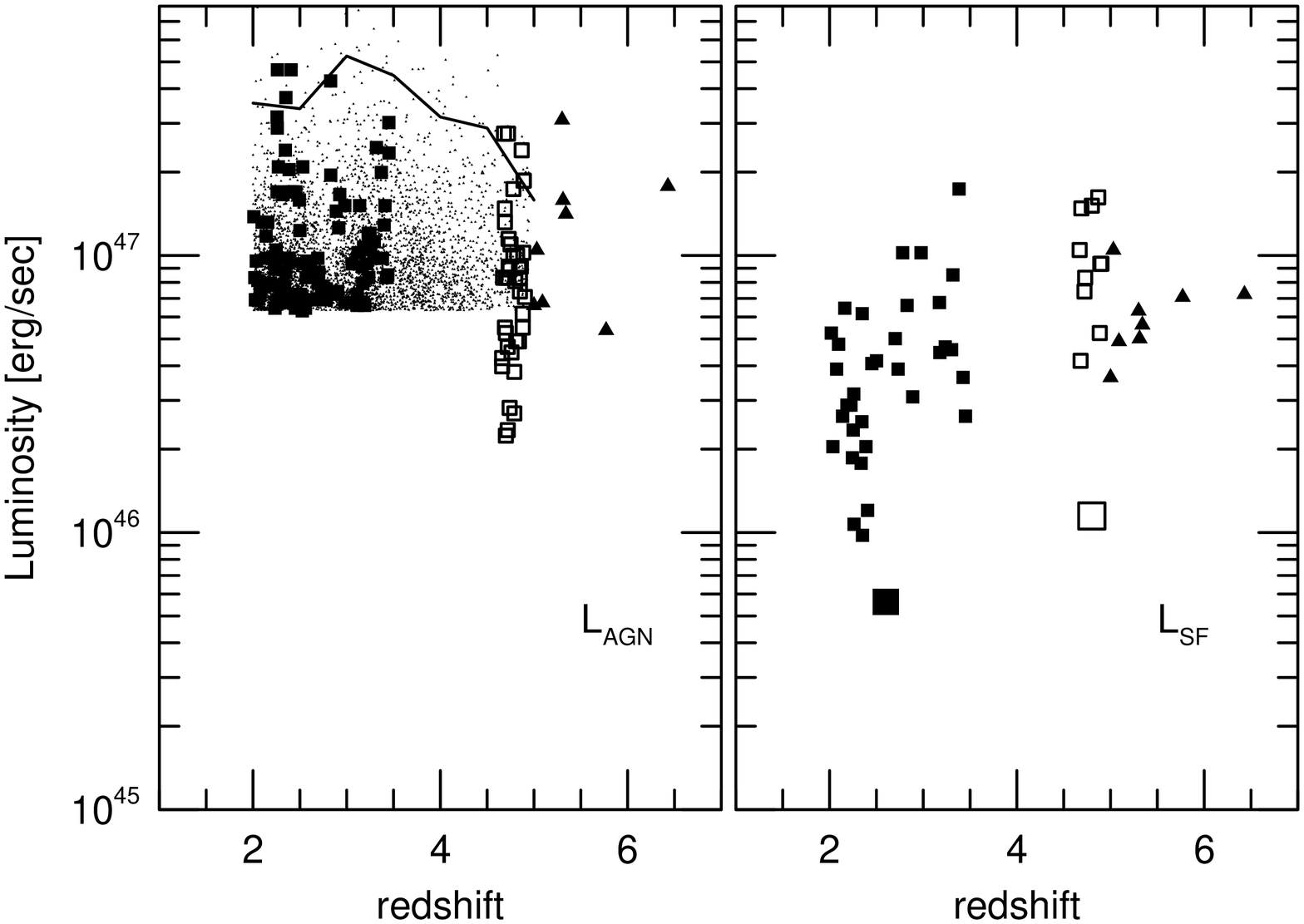}
  \caption{\LAGN\ (left panel) and \LSF\ (right panel) as functions of redshift for \herschel-observed sources in the \cite{Leipski2014} (filled
triangles, detections only),
\cite{Netzer2014} (small open squares for detections and a large open square for the stack),  
the present samples (small filled squares for detections and large filled squares for the stacks), and the parent SDSS sample (small points).
The solid line in the left panel connects the 10th most luminous AGNs in the \cite{Shen2011} sample  in redshift bins separated by 0.5.
}
  \label{fig:LAGN_LSF_z}
 \end{figure}

Fig.~\ref{fig:LAGN_LSF_z} shows the changes with redshift of   
\LAGN\ and \LSF\ for all the detected sources in
the three high redshift samples, and the stacks in the current sample and in the \zzzz\ sample.
While the \cite{Leipski2014} sample is less complete, and we do not have information on the undetected sources, it allows us to extend
the redshift range to beyond 5.
The dependence of the highest \LAGN\ on redshift is not new. It is similar to what is known from previous studies of large samples 
like the SDSS, and from the X-ray and optical LFs of AGNs
 \cite[e.g.][]{Croom2009,Vestergaard2009,Ueda2014,Vito2014,Shen2011,Trakhtenbrot2012}
(see however a different view in \cite{Vardanyan2014} regarding the most luminous AGNs based on \wise\ observations). 
As already shown here, and in \cite{Netzer2014}, both the \zzzz\ and \zzz\ samples 
represent the population very well and thus reproduce the behavior of the highest luminosity part of the LF
This is illustrated also by the entire parent SDSS sample shown in the left panel as small points.
The left panel of Fig.~\ref{fig:LAGN_LSF_z} shows a steady increase from very high redshifts up to
$z\sim 3$ followed by a decrease at smaller redshifts. This trend is illustrated 
in a simplistic way by showing a line representing the 10th most luminous source from the  \cite{Shen2011} catalog
(this way of representation is superior to the median \LAGN\ that reflects only the chosen lower limit on \LAGN\ because of the very steep 
LF).

The changes of \LSF\ with redshift for the hosts of the most luminous AGNs, shown in the right panel, is new
and very different. It exhibits a moderate rise from $z=7$ to $z\sim 4-5$ followed by a 
decrease at lower redshifts.
 The diagram shows a hint for an overall decline in \LSF\ from z=3.5 to z=2.
 Unfortunately, the statistics is rather poor and based on only 34 \herschel-detected sources.
However, it is in general agreement with studies of IRLFs such as \cite{Gruppioni2013}. Such studies show a steady increase in
\LSF\ for the most luminous FIR galaxies from $z=2$ to $z\sim 4-5$ with a rather uncertain behavior beyond this redshift due to poor statistics.
Moreover, the luminosity 
close to the high luminosity end of the $z \sim 4-5$ IRLF
 ($\log$~\LSF\ (\ergs)$\sim 47$) is very similar to the luminosities measured by us at the same redshifts.
We can thus conclude that the host galaxies of the most luminous AGNs across the redshift range $z=2-5$ represent well the high luminosity end of the IRLFs at those redshifts. We note also a somewhat similar behavior, based on a smaller number of sources, in the type-II radio-loud sample of 
\cite{Drouart2014}.
In summary, the analysis of the most luminous AGNs and their hosts shows that the peak in BH growth rate lags behind the peak
SFR of their host galaxies by approximately a Gyr, corresponding to the difference in cosmic time between redshifts $\sim5$ and $\sim3 $
under the assumption that we are looking at the same population in these redshifts bins..

Our current sample does not contain a large enough number of AGNs with reliable BH mass measurements and using such information for only
a handful of sources, and not others, can result in sever biases. We prefer to use estimates of these properties
in earlier studies of the highest luminosity AGNs in the same redshift range. The most suitable studies are those of 
\cite{Shemmer2004} and \cite{Netzer2007a} with a combined number of 44 AGN between z=2 and z=3.5. The luminosity distribution in 
this sample is different
from the one used here, since it contains several AGNs below our lowest \LAGN, and several others that are even more luminous than the SDSS 
sources studied here.
However, the median $\log$(\LAGN) is 46.9, merely 0.1 dex below the median value in the present sample. For a lack of better information, we assume the
same median Eddington ratio for the two samples and hence adopt the median BH mass in the above combined sample, $10^{9.5}$~\msun, as our best estimate for the median BH mass 
in the current sample\footnote{Note that the Eddington ratio in this sample is considerably below 1 while the typical values
at \zzzz\ are significantly higher}.
The situation at \zzzz\ is better since all objects have reliable BH mass estimates.\footnote{The median BH mass for the \herschel-detected sources at $z \approx 4.8$ is $10^{8.9}$~\msun.} 

Unfortunately, we do not have stellar mass measurements for any of the sources in the \zzz\ and \zzzz\
samples. We therefore limit the discussion
to a simple scenario where the stellar mass is not very different
from the largest MS stellar masses at those redshifts. This is in the range $10^{11-11.5}$~\msun\ \citep{Schreiber2015}. 
for the \zzzz\ sample and perhaps somewhat higher at \zzz.
Such a scenario is one likely possibility out of several that we are not yet in a position to test.
We are going to test the consequences to 
BH and stellar mass growth across the redshift range of \zzzz\ to \zzz. Such ideas are rather speculative given all the unknowns mentioned
above.

Being the largest mass objects, it is reasonable to assume
that the objects observed in the two redshift intervals represent  
the population that will eventually become the most massive BHs in the most massive galaxies in the local universe. 
As explained in \cite{Netzer2014}, this is a likely but by no means the only possibility.
\cite{Netzer2014} considered various possibilities regarding the location of the host galaxies of the most luminous AGNs at \zzzz\ 
relative to the MS at that redshift. The discussion included various scenarios such as strong and moderate feedback, 
the accumulation of stellar and BH mass under
various growth modes (exponential and linear growth), and the required duty cycles to explain the observations available at that time. 
We do not repeat this analysis partly because all the details are given in \cite{Netzer2014}, and partly
because of the lack of BH mass measurements at \zzz. Instead we report, briefly, on those
scenarios that are consistent with the new observations and point out some tests that can be done to verify
these ideas.

For simplicity, we consider BH growth during 1Gyr between \zzzz\ and $z\simeq 2.9$, taking the latter to represent our \zzz\ sample
(the median redshift of our current sample is somewhat lower, about 2.6).
 The mean \LAGN\ over this period translates to BHAR$\sim 8$~\msunyr. Therefore, 
accumulating the additional BH mass (an increase by a factor 4 growing from $10^{8.9}$ to $10^{9.5}$ \msun) requires a linear growth with a duty cycle of approximately 0.5. 
For exponential growth, the duty cycle is, of course shorter.

For the stellar mass growth, we assume that at \zzzz\ the host galaxy mass is in the range $10^{11-11.5}$~\msun. According to 
 \cite{Schreiber2015}, at \zzzz\, 
the SFR for a $10^{11}$~\msun\ MS galaxy
is about 260 \msunyr\ and for a $10^{11.5}$~\msun\ MS galaxy, about 790 \msunyr. The corresponding numbers at $z\simeq 2.9$ are about 140 and 330 \msunyr, respectively.
Assuming we are following the same population in time, it is reasonable to suggest that the lowest stellar mass to consider at \zzz\ is $10^{11.5}$~\msun.
Using these stellar mass estimates, and assuming a typical MS width of $\pm 0.3$ dex, we suggest that most of the \herschel-detected sources 
at \zzzz\ are above the MS and most of the undetected sources at that redshift are on the MS.
As for the \zzz\ sample, here the
median SFR for the stacked sources
is roughly 100 \msunyr\ (Table~4) which means that most, or perhaps all objects with stellar mass of $10^{11.5}$~\msun\ or larger  are below the MS. 
This is consistent with the assumption that all the \herschel-undetected sources at \zzz, i.e. about 2/3 of the objects in our sample, are
quenching or possibly quenched galaxies. Obviously the distribution in SFR for undetected sources is likely to be wide and the information that
we have is only about the mean and medians properties of the population.
We did not consider SF hosts with stellar masses smaller that $10^{11}$ \msun\ at \zzz\  
which is probably unphysical given the very large BH mass expected in these very luminous AGNs.

Given all these numbers, a stellar mass growth at a rate of $10^3$~\msunyr\ starting at \zzzz, gives just
enough time to accumulate mass which, at \zzz, is similar to the mass of the most massive galaxies of today. 
The estimate is basically independent of the starting stellar mass.
This scenario is consistent with  known stellar mass functions at the same redshift
range \citep{Ilbert2013,Madau2014} provided 
a complete quenching occurs towards the end of this redshift interval.
Faster stellar mass growth, which is 
more consistent with the sSFR of the \herschel-detected sources at \zzzz, requires less time, with a duty cycle
of the order of 0.2, or quenching at redshift larger than 3.5, consistent with ideas about the epoch of fastest growth of the
most massive galaxies of today. 

Regarding undetected sources, and assuming again we observe the same population at both redshifts,
most such objects are still on the MS at \zzzz\ but below the MS at \zzz.
Either the hosts of these BHs are low mass SF galaxies, which is problematic given the assumed very large BHs at \zzzz, or else
quenching becomes important between these two epochs.
If the latter is correct, then the active BHs we observe in this group are consuming the remaining gas near the center while SF in the galaxy 
has ceased.
These BHs may be on their way to become the most massive BH known. A similar phase may occur at redshift smaller than 2 for those sources that are
still forming stars at high rate at \zzz.

It is interesting to note that the scenario described above for \herschel-detected sources at \zzzz\ and \zzz, 
that identify them as the most massive systems
of today, is consistent also with the measured SFR/BHAR.
In  most of these sources, \LSF$\sim$\LAGN. For this luminosity ratio,  
 SFR/BHAR$\sim 140$ \citep{Netzer2014}, a ratio which is very close to the stellar-to-BH mass ratio observed in the spheroidal galaxies 
hosting the most massive BHs at z=0.

Finally, the suggestion that the \herschel-detected sources at \zzzz\ and \zzz\ represent the same population, put a question mark
on the idea that AGN feedback in such sources is an important  process that regulates their stellar mass growth. Such 
extremely luminous AGNs are active, for a long period of time, without affecting much the very fast SF in their host galaxies.

\subsection{Torus properties: SED and covering factor}
\label{sec:torus_properties}

\subsubsection{Constraints on the torus SED}

The observations presented here can be used to obtain information about several of the nuclear components in the objects under study. In particular, 
we can set limits on the shape of the torus SED in highly luminous sources, thus improving the estimates of the total dust
emission by the torus, and estimate more accurately its covering factor (\Cf). We can also 
look for signs of intrinsic reddening due to interstellar or circumnuclear dust in the host galaxy. 

Constraints on the short wavelength part of the torus spectrum 
can be obtained from study of the data presented in Fig.~\ref{fig:stack_L1}. Interestingly, detected and undetected \herschel\ sources
display remarkably similar shapes over the 1--10 \mic\ range that are all consistent with the median SED of \cite{Mor2012}. 
The long-wavelength part of the 
torus spectrum is more difficult to observe due to contamination by SF in the host galaxy. This part is best studied by using SPIRE upper limits. 
We tested several different suggested torus SEDs.  The first is the \cite{Mor2012} template used throughout this paper. 
Here the turning down (in $\lambda L_{\lambda}$) is at wavelengths greater than $\sim 25$~\mic. 
The second is a set of three SEDs published by \cite{Mullaney2011} (listed
in their Table 3). All three are flat, in $\lambda L_{\lambda}$, up to 30--40\mic\ and drops down, in a luminosity dependent way,
beyond this wavelength. The final SED is the one used 
by \cite{Tsai2015} partly in attempt to fit the global spectrum of highly obscured, luminous AGNs. This template was 
adapted from the earlier works of \cite{Polletta2006} and \cite{Polletta2007}. It is similar to the \cite{Mor2012} SEDs
at short wavelengths but extends to longer wavelengths with a drop starting at around 50\mic.
The main difference between the \cite{Tsai2015} SED compared to the \cite{Mor2012} and \cite{Mullaney2011} SEDs is
that the former do not take into account the SF contribution at FIR wavelengths."

\begin{figure}[t]
\centering
\includegraphics[height=6cm]{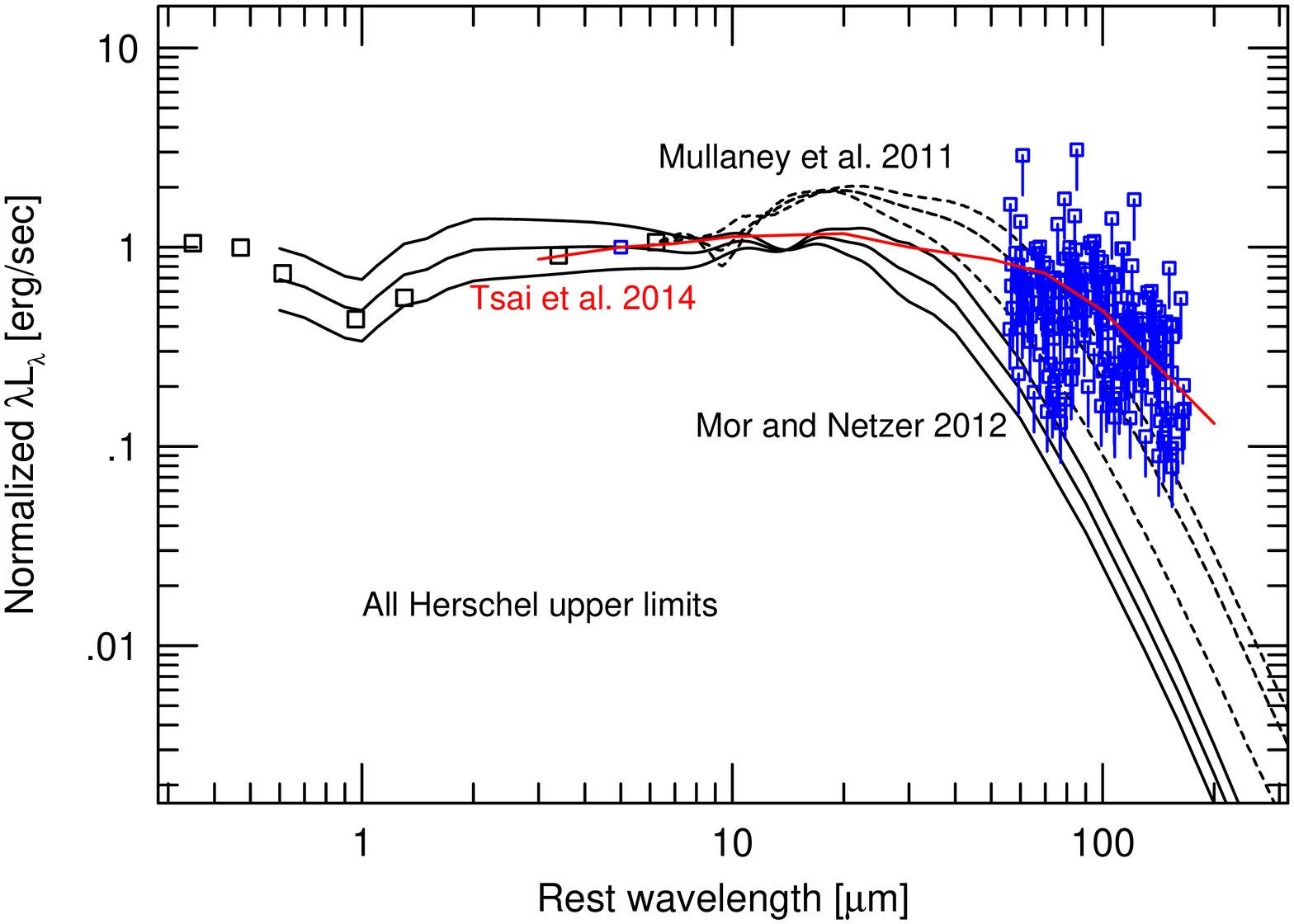}
\includegraphics[height=6cm]{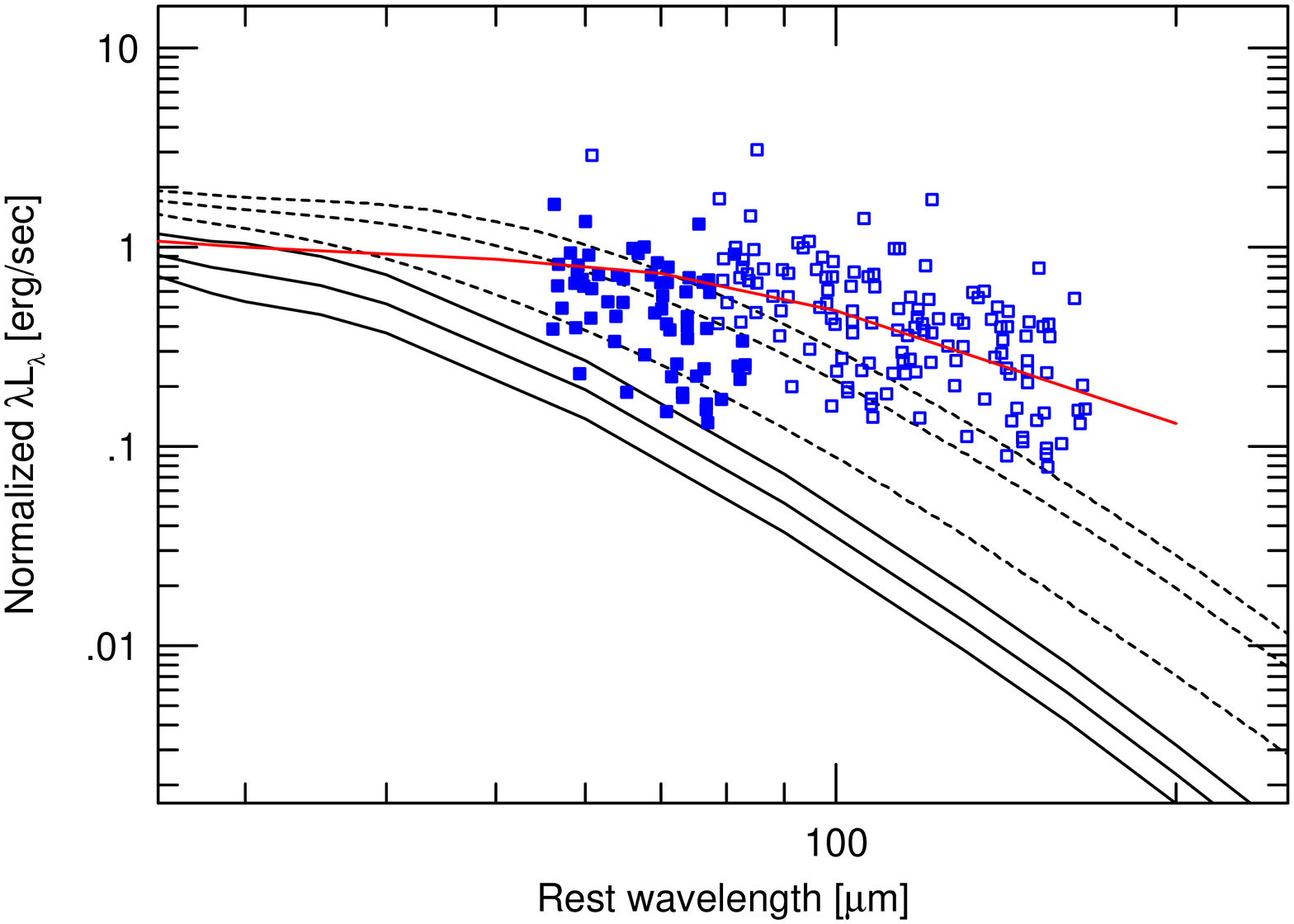}
  \caption{Left: The shape of the torus SED from FIR upper limits. 
The diagram shows all \herschel/SPIRE upper limits for the undetected sources in our sample normalized to \Ldust\ derived from the
\WISE\ data. The short wavelength points show the NIR-\WISE\ data of a representative object  
(HE-2156-4020 at z=2.531). The central solid black line is the \cite{Mor2012} SED used throughout the paper and the two other solid lines the 25\% and 75\%
limits discussed in that paper. The dashed black lines are the three SEDs suggested by \cite{Mullaney2011} 
normalized to the \cite{Mor2012} SED at 6~\mic: 
from top to bottom, high-L, mean and low-L sources.
  The SPIRE upper limits indicate that if all sources have the same SED, they 
  cannot exceed the \cite{Mor2012} template  by more than approximately  10\% at rest-frame wavelengths of 60--90 \mic. 
The red line is the \cite{Tsai2015} torus SED normalized to the observed 5~\mic\ continuum. 
Right: zoom in on the long wavelength part (error bars removed for clarity). The full squares mark the 250~\mic\ upper limits.
 }
  \label{fig:torus_upper_limit}
 \end{figure}

The stack spectrum shown in Fig.~\ref{fig:stack_L1} provides strong constraints on the torus SED and demonstrates that it cannot 
exceed the \cite{Mor2012} SED 
by a large amount at rest-wavelengths of 60-90\mic\ since the flux emitted by the torus cannot exceed the stacked fluxes.
Individual SPIRE upper limits (i.e. 3 times the confusion limits) provide even stronger constraints.
This is illustrated in Fig.~\ref{fig:torus_upper_limit} that shows the \cite{Mor2012} SEDs (composite and upper and lower limits), 
the three \cite{Mullaney2011} SEDs,  and the \cite{Tsai2015} SED all
normalized to our median observed SED at 5--6~\mic. 
The left panel of the  diagram shows $3 \sigma$ upper limits in the three SPIRE bands for the 66 undetected sources in our sample and
the right panel is a zoom in on the more important part of the diagram. 
While we cannot exclude the possibility that the torus SED changes from one source to the next, we can put strong
constraints on its shape under two simplified assumptions. The first assumption is that 
all sources have the same torus SED within the 25--75\% range set by \cite{Mor2012}. 
The diagram shows that all our sources are consistent 
with this assumption. The high-L SED of \cite{Mullaney2011} is consistent with most but not all objects.

Alternatively, we can test the assumption that the \cite{Mullaney2011} SEDs, or the  \cite{Tsai2015} SED, provide good fits to
the undetected sources by comparing their predicted 250/(1+z)~\mic\ luminosity to the observations.
For the high-L \cite{Mullaney2011} SED, we find that 24\% of the undetected \herschel\ sources would have been detected by SPIRE  
at a 3$\sigma$ level if this was, indeed, the torus SED. For the mean \cite{Mullaney2011} SED, this number is 71\% and for the 
\cite{Tsai2015} SED, 85\%. 
These tests suggest that the \cite{Mor2012} SED is the best choice for our sample and and some of the alternative SEDs tried here
are less consistent with the observations. 
This justifies our earlier assumption that 
except for the weak-NIR sources, simple integration over the chosen SED is appropriate for  estimating
the torus covering factor in our sample.

\subsubsection{Intrinsic reddening}

We also considered the possibility of intrinsic reddening in our sources. For this we need to
compare the estimated \LAGN, which can be affected by reddening, and \LTOR, 
which is independent of reddening. Our comparison is based, again, 
on the torus SED adopted here which fits well all the sources except for the weak-NIR AGNs, where it clearly
overestimates \LTOR. 
We have 12 weak-NIR sources
that represent 12\% of our sample, very similar to their fraction in the general population
\cite[e.g.][]{Mor2011,Roseboom2013}. The correlations and diagrams 
described below do not include these sources\footnote{Both \cite{Roseboom2013} and
\cite{Leipski2014} use individual torus model for every source and hence their measured \LTOR\ represent well the total dust emission
by the torus.}.

Intrinsic reddening is hard to check in individual
sources because of the large scatter in the intrinsic shape of the optical-UV continuum
\cite[e.g.][and references therein]{Krawczyk2013,Krawczyk2015}.
\cite{Lusso2013} discussed $z=0-5$ AGNs and assumed a single disk-like SED adopted from an observed 
composite by \cite{Richards2006}. The normalization of the SED is based on the multi-band photometry of their sources.
They find that 24\% of the sources in their sample are affected by significant reddening corresponding to $\langle E(B-V) \rangle = 0.1$ mag., 
where the sample mean is $\langle E(B-V) \rangle =0.03$ mag.
A major limitation of this method is the assumption that the SED of the accretion disk, assumed to be the central power-house,
 is independent of the source luminosity and BH mass, and the bolometric luminosity is independent of the disk inclination to the 
line of sight.  This assumption is in contrast with calculated thin disk SEDs 
that depend on
BH mass, BH accretion rate, and BH spin \cite[e.g.][and references therein]{Capellupo2015}.

We chose not to make specific assumptions about the origin of the intrinsic AGN SED but rather to look for significant
variations in continuum slope as reddening indicators. 
For this we compared the rest-frame \LUV, and the luminosity derived from the K-band flux which is available for about half the sources
in our sample. We used these measurements to derive the continuum slope, $\alpha$, between the two wavelengths. The K-band central wavelength  
corresponds to rest-frame wavelengths between 0.49 and 0.73\mic, depending on the redshift. Given the rough nature of this comparison, we
 did not take into account the contribution to the K-band flux from H$\beta$ and FeII lines in objects with 
$z > 3$, and the contribution from H$\alpha$ in sources at $z\sim 2$.
The distribution in $\alpha$ calculated in this way, assuming
$L_{\nu} \propto \nu^{-\alpha}$, is broad with $\alpha$ in the range 0--1, and resembles the general distribution in large samples of 
unreddened AGNs
\citep{Krawczyk2015}. The median slope for the entire sample can be found by comparing the median IR-SED, extended  to below 1\mic, and the
median \LUV. This slope is $\alpha=0.22$, again similar to what is found in large AGN samples.
We find no correlation of slope with \Ldust/\LUV\ and, therefore, neglect the
effect of intrinsic reddening on \LTOR/\LAGN\ in our sample. Note, again, that such information is only available for about half the sources in our sample.

\begin{figure}[t] 
\centering
\includegraphics[height=8cm]{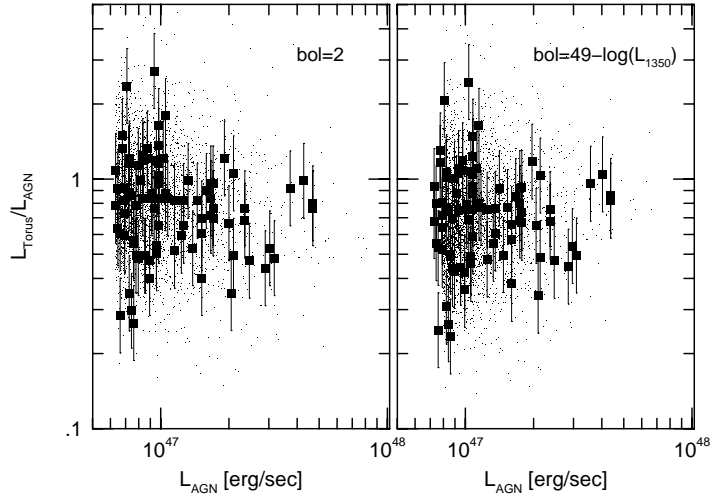}
  \caption{\LTOR/\LAGN\ for the \zzz\ sample excluding weak-NIR sources (large symbols), and the comparable \cite{Shen2011} sample
 (small points)).
The two panels illustrate the changes resulting from the use of different bolometric correction factors (marked in the diagram as ``bol'').
  }
  \label{fig:covering_factor_1}
 \end{figure}

Fig.~\ref{fig:covering_factor_1} shows \LTOR/\LAGN\ as a function of \LAGN\ for all the sources in  our sample except for the
12 weak-NIR sources whose \LTOR\ is more uncertain. 
We calculated the ratio under the two different assumptions considered earlier for the bolometric correction factor, $Bol_{1350}=2$ in the left panel
and $Bol_{1350}=(49-\log$\LUV) in the right panel. A regression analysis using the BCES method gives, in the first case, a marginally significant
slope of $-0.46 \pm 0.27$, and a slope which is consistent with no correlation ($-0.35 \pm 0.86 $) in the second case. Thus
the uncertainty in \LAGN\ due to the large possible range in $Bol_{1350}$) completely masks any real dependence of \LTOR/\LAGN\ on \LAGN.
We also show (small black points) similar data for the partial \cite{Shen2011} $\log$~\LUV\ (\ergs)$>46.5$ sample discussed above in \S~\ref{sec:torus_sed}.
The dependence of \LTOR/\LAGN\ on \LAGN\ is similar and the conclusions are unchanged.
Below we investigate this issue using results from several other samples.

\subsubsection{Covering factor}
\label{sec:covering_factor}

Next we consider the geometrical covering factor of the torus, \Cf. For this we need 
to consider possible anisotropies in \LTOR,  \LAGN, and their ratio. 
The first source of anisotropy is the geometry of the central power-house considered here to be optically thick, geometrically
thin or slim accretion disk. For thin disks, 
the angular dependence of the emitted radiation, neglecting general relativistic effects at very high frequencies,
 is wavelength independent and varies roughly as $\cos i$, or $\cos i (1+a \cos i )$, where $i$ is the inclination angle
and $a \approx 2$ 
\cite[e.g.][]{Netzer2013}.  For a thin disk whose axis is parallel to that of the torus, in a type-I AGN, this factor
is in the range 0.5--1, representing the range of inclinations from zero to 60 degrees. The expected anisotropy in slim accretion disks 
is much larger \cite[][and references threin]{Wang2014b}.
The calculations of the SED, in this case, are far more complicated, and therefore highly uncertain.  

The second sources of anisotropy is the dusty torus itself. The radiation pattern of such tori have been discussed, extensively, 
in the literature  \cite[e.g.][and references therein]{Nenkova2008b,Stalevski2012} and reviewed recently
by \cite{Netzer2015}. Strong anisotropy, especially at short wavelength where the dust optical depth is the largest, 
is predicted by most torus models. The exact
angular dependence differs substantially from one model to the next, partly because of the different geometries used in such calculations. 
For example, a new work by Stalevski (2105; private communication) suggests
 that for both continuous and composite clumpy tori,
the torus is very difficult to detect for very small \Cf, because of the inner disk anisotropy, and the relation between 
\LTOR/\LAGN\ and \Cf\ is  non-linear at large covering factors. 

Given the large uncertainties, we decided to 
adopt the simple anisotropy correction factor of \cite{Netzer2015} which is similar to those used by \cite{Treister2008} and \cite{Lusso2013}
(\cite{Roseboom2013} did not apply an anisotropy correction factor and all their results refer to the case of complete isotropy).
For this we introduce an isotropy parameter, $b$, that can vary between 1 (complete isotropy, i.e. the part of the radiation emitted into the torus opening is proportional
to the solid angle of the opening in the torus) and 0 (complete anisotropy, 
all torus emission is emitted into the opening).
In this case,
\begin{equation}
\frac{ L_{\rm torus}}{ L_{\rm AGN}} =  \frac{ 1 - b C_f }{1-C_f}  C_f  \, .
\label{eq:covering}
\end{equation}
This expression does not take into account anisotropic disk emission that affects \LAGN\ and is equivalent to the assumptions of the same disk inclination
angles in all type-I AGNs. It does not necessarily increase the uncertainty since if the disk and the torus axes are aligned, the ration \LTOR/\LAGN\ depends
less on inclination to the line of sight.

The parameter $b$ depends on the optical depth of the dust in the torus and is therefore wavelength dependent. The dependence affects the conversion between the observed
\LTOR\ and the total dust emission, and also the scaling of the total dust emission relative to the measured \Ldust\ (i.e. the factor of 3.58 
introduced in \S~\ref{sec:torus_sed}).
 For the extreme cases of very small optical depth (complete isotropy), or large optical depth over the entire 2--20\mic\ range, this dependence will not
affect the derived covering factor.  Here we neglect the wavelength dependence of $b$.

We compiled from the literature a large number of estimated covering factors based on NIR-MIR observations. They
include:
 1) The COSMOS sample of \cite{Lusso2013}. Here we take median data shown in Fig. 12 of their paper.
2) The \cite{Roseboom2013} \WISE-based $z \leq 1.5$ sample. Here we take the mean and standard deviations from their Fig.4 but made an adjustment to correct for the
fact that their bolometric correction factors obtained from \cite{Shen2011} are significantly larger than those in our
work  (\S~\ref{sec:bolometric}). 
This adjustment results in a
considerable increase in the mean \LTOR/\LAGN.
3). The \cite{Mor2012} sample. This sample includes $\sim 100$ low-to-intermediate luminosity objects collected from the literature including many 
narrow line Seyfert 1 galaxies and QUEST QSOs.
We used the observed \Lop\ to obtain \LAGN\ and \Ldust\ to obtain \LTOR. We did not use the estimated covering factors listed in their paper that are based
on the specific anisotropy provide by the \cite{Nenkova2008a} model. This is a completely different, model dependent way which is different from the method used
for obtaining \Cf\ in the other samples.

Several other samples use NIR-MIR observations to calculate covering factors. 
 In particular, \cite{Treister2008} present covering factor estimates based on flux measured in a single \spitzer\ band (24\mic). They made 
several assumptions about the total torus emission and calculated \LAGN\ in a way which is somewhat different from what we used here. Intriguingly, 
their estimated \LTOR/\LAGN\ is significantly larger than what was obtained by others for a similar range of luminosity and redshift.
The discrepancy was discussed by \cite{Lusso2013} without a real resolution. We decided not to consider this sample since the redshift and luminosity
ranges are nicely covered by other samples where \LTOR\ is better defined.
The \cite{Maiolino2007} multi-redshift sample used \spitzer-IRS data to constrain the torus SED.
This sample was discussed in great detail by \cite{Lusso2013} who showed a very good agreement with their results. Thus the data we present
here represent well the results of \cite{Maiolino2007}.

\begin{figure}
\centering
\includegraphics[height=8cm]{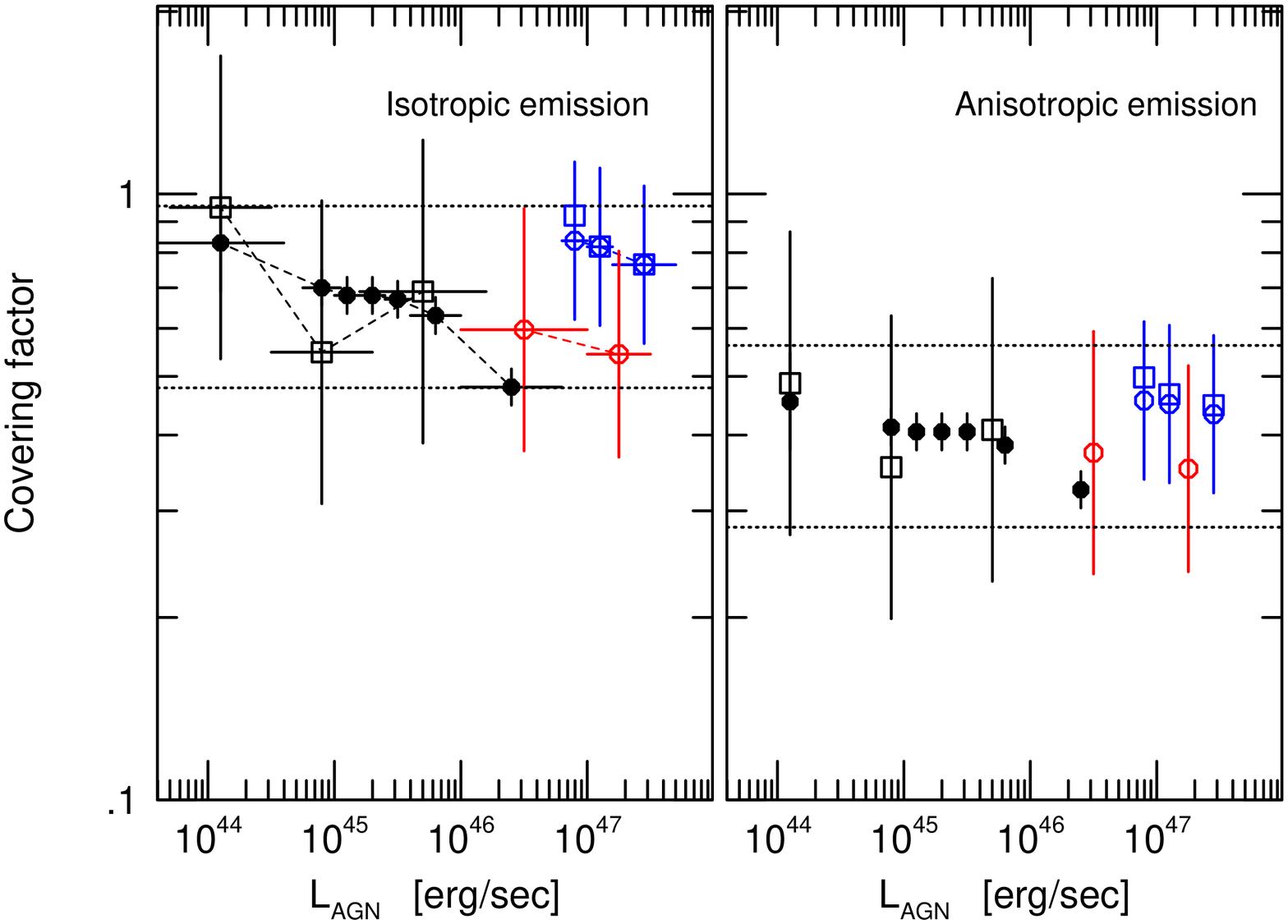}
  \caption{Covering factor as a function of \LAGN\ in large AGN samples. Open black squares: the \cite{Mor2012} sample.
Full black circles: the \cite{Lusso2013} sample. Open red circles: the \cite{Roseboom2013} sample. Open blue circles: the
the present $z=2-3.5$ sample without the weak-NIR sources. Open blue squares: SDSS sources from the \cite{Shen2011} catalog
with the same redshift and luminosity as the present sample and \Ldust\ measured as explained in the text.
The left panel shows the covering factors under the assumption of isotropic dust emission. 
As explained in the text, the \cite{Roseboom2013} values
are scaled up by factors of 1.5--1.8 to allow for the different assumed bolometric correction factors. 
The parallel dotted lines show the range of $\pm0.15$ dex around the mean value of 0.68.
The right panel shows the same data for the case of complete anisotropy ($b=0$ 
in eqn.~\ref{eq:covering}). The dotted lines indicate a range of $\pm 0.15$ dex around 0.4. 
  }
  \label{fig:covering_factor_2}
 \end{figure}

Fig.~\ref{fig:covering_factor_2} shows all the NIR-MIR-based compilations of \Cf, including ours, for 
the case of complete isotropic dust emission ($b=1$) on the left, and maximum anisotropy ($b=0$) on the right. 
For the present sample we only show the case of $Bol_{1350}=2$.
The uncertainty on
\LAGN\ represents the range in this property used to derive the median values. The uncertainty on \LTOR/\LAGN\ is  taken from the
original papers \citep{Roseboom2013,Lusso2013} or from the present calculations. Given the uncertainties on \LTOR\ and \LAGN, we estimate a 
combined uncertainty on \Cf\ of at least $\pm 0.2$ dex at all \LAGN. We also show, as blue open squares, the median values obtained from the \cite{Shen2011}
 and shown in Fig.~\ref{fig:covering_factor_1} left panel ($Bol_{1350}=2$).

The diagrams presented here show that all medians of all the samples used here, at all \LAGN,  are confined to a band of width $\pm 0.15$ dex in \LTOR/\LAGN\
around  0.68 for the isotropic case, and 0.4 for the case of complete anisotropy.
Since much of the uncertainty in \Cf\ is systematic, and depends on the poorly known bolometric correction factor, 
the inclusion of of a large number of individual measurements
cannot improve the situation by much.
Thus, torus covering factors which are based on NIR-MIR measurements, show no indication for a decrease of \Cf\ with increasing source luminosity.
This finding seems to be in contradiction with earlier findings such as those presented by \cite{Lusso2013} and \cite{Roseboom2013}. 

An independent way of estimating the covering factor is to compare the fraction of type-II AGNs in the AGN population as a function of redshift
and luminosity either by searching for X-ray obscuration or by counting sources in large optical surveys. 
This has been an active area of research for many years and was investigated in numerous X-ray papers 
\cite[e.g.][]{Steffen2003,Ikeda2009,Yaqoob2010,Brightman2011a,Ricci2013,Buchner2015}. A additional way is to compare
type-I and type-IIs' LFs \citep{Simpson2005}. A detailed comparison of X-ray derived and IR-derived covering
factors is beyond the scope of the present paper. We only point out that disagreements between the various methods have been noted in earlier works
especially for low luminosity AGNs, and are partly due to the different cut-offs in column density adopted in different papers
 \cite[see a comprehensive discussion in a recent paper by][]{Merloni2014}. There are still fundamental unresolved issues related to the relative
fraction of type-I and type-II sources at different luminosity and redshift. One such difficulty is related to the 
``true'' type-II AGNs found in low luminosity samples. Another uncertainty is related to the difficult-to-detect broad wings in low luminosity 
type-I sources \citep{Oh2015}. 
A good example of a difficulty to asses the covering factor in X-ray samples is the recent work by \cite{Vito2014}  which classifies
 AGNs into groups based on line-of-sight absorbing column of $10^{23}\,{\rm cm}^{-2}$, which is considerably larger than the column densities 
used in earlier studies. This work suggests a roughly 1:1 ratio
of type-I and type-II AGNs over a large range of X-ray luminosity which overlaps, given standard bolometric correction factors, 
with the luminosities in our sample.  This would indicate \Cf$\sim 0.5$.
A comprehensive review of many of these issues in given in \cite{Netzer2015}.

The above finding question the validity of the ``receding torus'' model suggested by \cite{Lawrence1991} and discussed in numerous
other papers. The model aims at explaining the seemingly decreasing covering factor of AGN tori as a function of \LAGN.
It is based on observational and theoretical ideas that the innermost boundary of the central dusty torus is defined by the dust sublimation
radius \cite[e.g.][]{Barvainis1987,Netzer1993}. Recent studies \cite[][and references therein]{Koshida2014} based on dust reverberation mapping, 
show the good
agreement between the dust innermost location and the sublimation radius of pure graphite dust \citep{Mor2012}. In particular, they show a clear 
dependence of the the innermost dust
location on \LAGN$^{1/2}$, as expected in the simplest model of this type. 
The receding torus idea takes this idea one step further by assuming  that the vertical scale of the torus (the torus ``height")
is independent, or only weakly dependent, on source luminosity. 
This assumption results in a smaller covering factor for larger \LAGN.
There is little if any theoretical justification of this idea.

The present work suggests that, given the uncertainties, the covering factors of tori in the 
most luminous AGNs may be very similar
to those in sources that are three orders of magnitude less luminous. For the simple torus models this is equivalent to a factor of $\sim 30$ in distance between
the central BH and the torus inner walls.
We suggest that earlier claims to the contrary could be biased mostly by the inconsistent 
use of various bolometric correction factors and that the overall geometry (shape and size) of AGN tori scale in accord with the bolometric luminosity.
The evidence presented here is based on intermediate to high luminosity AGNs (see luminosity scale in Fig.~\ref{fig:covering_factor_2}) and hence does 
 not apply to lower luminosity AGNs in the local universe.
A future, more detailed study
of the covering factor distribution as a function of \LAGN\ must be limited to a narrow redshift range to avoid evolutionary biases.

\section{Conclusions}

The \herschel/SPIRE observations reported here provide new information about 100 very luminous, optically selected type-I AGNs 
at \zzz\ with $\log$~\LAGN\ (\ergs)$\ge 46.8$, assuming \LAGN=2\LUV.
Our sample provides the most complete information, in terms of numbers, about this population since there are very few such objects in other 
\herschel-selected fields. The distributions in \LUV\ and \Ldust\ of the sources is similar to the distributions in the general
population (SDSS), and we can use the sample to study several outstanding problems related to SF and BH activity in such sources. 
In particular, we can
combine the sample with two previous studies at high redshift, those of \cite{Netzer2014} and \cite{Leipski2014},
 and follow BHAR and SFR, and their ratio, over the redshift interval 2--7, albeit with incomplete
information for the $z> 5$ population. The main results are:
\begin{enumerate}
\item
Of the 100 sources, 34 are detected by \herschel\ at the $3\sigma $ level. For the undetected sources, we
present two statistically significant stacks representing sources in two luminosity groups: $\log$~\LUV\ (\ergs)=46.5-46.7 and  $\log$~\LUV\ (\ergs)$>$46.7.
The mean and the median SFRs of the detected sources are $1176^{+476}_{-339}$ and $1010^{+706}_{-503}$ \msunyr,
respectively. The mean SFR of the undetected sources is $148$ \msunyr (uncertainty not given since we did not stack the entire group of undetected sources; see
Table~4 for more information)..
Unlike our earlier $z\simeq 4.8$ sample, the \zzz\ sources do not show significant differences in \LAGN\ and \LTOR\ between 
\herschel-detected and undetected sources.
\item
The combination of the three high redshift samples show that the redshift distribution of \LSF\ and \LAGN\ for the most luminous, 
redshift 2--7 AGNs are different. 
 Like the entire SDSS sample, the highest \LAGN\
increases with decreasing redshift, peaking at $z\approx 3$. However, the highest \LSF\ in the host galaxies of the most luminous AGNs
increases with decreasing redshift more rapidly
and peaks at $z \approx 5$.
 Assuming the objects in our sample are hosted
by the most massive galaxies at \zzz, we argue that some 30\% of the hosts are on and above the MS and most of the remaining 
70\% are below the MS.
The ratio of the stellar to BH mass growth rate is $\approx 80$ in the high SFR, \herschel-detected sources, and less than 10 in the 
group of low SFR galaxies.
\item
The shapes of the SEDs of the dusty tori in our sample, as derived from a combination of \wise\ and J,H,K photometry,  are very similar to the shapes found  
in low redshift, low luminosity
AGNs. 
The measured \herschel\ upper limits put strong constraints on the long wavelength part of this SED.
The upper limits are in good agreement with the \cite{Mor2012} composite torus SED, 
in somewhat worse agreement with the high-L and mean \cite{Mullaney2011}
SEDs, and in contradiction with the \cite{Tsai2015} SED where the turning down is at very long wavelengths.
\item
Combining our results at \zzz\ with those of several earlier studies, and correcting for biases due to different bolometric correction factors 
used in the earlier works, we find no evidence for a luminosity dependence
of the torus covering factor in sources with 
$\log$~\LAGN\ (\ergs)=44-47.5. This conclusion is based on various
assumptions, mostly the recognition of the large uncertainties in several earlier calculations of \LAGN.
The median covering factors over this range are $0.68 $ for isotropic dust emission and $0.4 $          
for anisotropic emission, with an uncertainty of 0.15 dex on both numbers.
\end{enumerate}


\begin{acknowledgements}
We thank the anonymous referee for providing numerous suggestions that helped to improve the presentation of the paper.
This work is based on observations made with \herschel, 
a European Space Agency Cornerstone Mission with significant participation by NASA. 
The Herschel$-$ATLAS is a project with Herschel, which is an ESA space observatory with science instruments provided by European-led Principal Investigator consortia and with important participation from NASA. The H$-$ATLAS website is http://www.h-atlas.org/.
This research has made use of data from HerMES project ($http://hermes.sussex.ac.uk/$). HerMES is a $Herschel$ Key Programme utilizing Guaranteed Time from the SPIRE instrument team, ESAC scientists and a mission scientist.
The HerMES data was accessed through the $Herschel$ Database in Marseilles (HeDaM - $http://hedam.lam.fr$) operated by CeSAM and hosted by the Laboratoire d'Astrophysique de Marseilles.
Support for this work was provided by NASA through an award issued by JPL/Caltech. 
Funding for this work has  been provided by the Israel Science Foundation grant 284/13.
\end{acknowledgements}



\clearpage


\begin{sidewaystable}
\caption{The $z=2-3.5$ sample: positions, redshifts, UV and AGN luminosities. Right ascensions, declinations and UV luminosities are taken from \cite{Shen2011}; \LAGN ~was obtained as described in  \S~3.1.}
\vspace*{1cm}
\label{table_all_info1}
\tiny
\hspace{-2.5cm}
\resizebox{22.5cm}{!}{
\begin{tabular}{ l c c c c c c }
 \toprule
           $ID$ & $RA$ & $Dec $ & $z$ & $\log L_\text{1350}$  & $\log L_\text{AGN}$  & $Scan~ ID$ \\
              & (deg) & (deg) & & (erg/s) & (erg/s) \\ 
         \midrule
SDSS J002025.22$+$154054.7 & 5.1051 & 15.6819 & 2.0087 & 46.84 & 47.14 & 1342213198\\
SDSS J005202.40$+$010129.2 & 13.0100 & 1.0248 & 2.2706 & 47.02 & 47.32 & 1342201380\\
SDSS J005229.51$-$110309.9 & 13.1230 & $-$11.0528 & 2.4524 & 46.50 & 46.81 & 1342199390\\
SDSS J005814.31$+$011530.2 & 14.5597 & 1.2584 & 2.4949 & 46.90 & 47.20 & HerS\\
SDSS J010227.51$+$005136.8 & 15.6146 & 0.8602 & 2.5319 & 46.66 & 46.97 & HerS\\
SDSS J010612.21$+$001920.1 & 16.5509 & 0.3223 & 3.1196 & 46.71 & 47.01 & HerS\\
SDSS J011552.59$+$000601.0 & 18.9691 & 0.1003 & 3.1933 & 46.52 & 46.82 & HerS\\
SDSS J011827.99$-$005239.8 & 19.6166 & $-$0.8777 & 2.1861 & 46.60 & 46.90 & HerS\\
SDSS J012412.46$-$010049.8 & 21.0520 & $-$1.0138 & 2.8300 & 46.98 & 47.29 & HerS\\
SDSS J012517.14$-$001828.9 & 21.3214 & $-$0.3080 & 2.2780 & 46.59 & 46.89 & HerS\\
SDSS J012748.31$-$001333.0 & 21.9513 & $-$0.2259 & 2.0748 & 46.56 & 46.86 & HerS\\
SDSS J013014.30$-$000639.2 & 22.5596 & $-$0.1109 & 2.3847 & 46.69 & 46.99 & HerS\\
SDSS J013249.38$+$002627.1 & 23.2058 & 0.4409 & 3.1664 & 46.65 & 46.96 & HerS\\
SDSS J013654.33$-$003415.4 & 24.2264 & $-$0.5710 & 2.7317 & 46.56 & 46.87 & HerS\\
SDSS J014123.04$-$002422.0 & 25.3460 & $-$0.4061 & 2.5979 & 46.63 & 46.93 & HerS\\
SDSS J014214.75$+$002324.2 & 25.5615 & 0.3901 & 3.3704 & 47.00 & 47.30 & HerS\\
SDSS J014303.16$+$001039.6 & 25.7632 & 0.1777 & 2.5066 & 46.56 & 46.86 & HerS\\
SDSS J014733.58$+$000323.2 & 26.8899 & 0.0565 & 2.0400 & 46.53 & 46.84 & HerS\\
SDSS J014809.64$-$001017.8 & 27.0402 & $-$0.1716 & 2.1627 & 46.69 & 46.99 & HerS\\
SDSS J015017.71$+$002902.4 & 27.5738 & 0.4840 & 2.9774 & 46.53 & 46.83 & HerS\\
SDSS J015819.77$-$001222.0 & 29.5824 & $-$0.2061 & 3.3017 & 46.69 & 46.99 & HerS\\      
\end{tabular}}
\end{sidewaystable}

\newpage 

\addtocounter{table}{-1}
\begin{sidewaystable}
\caption{Continued.}
\vspace*{1cm}
\label{table_all_info2}
\tiny
\hspace{-2.5cm}
\resizebox{23.5cm}{!}{
\begin{tabular}{l c c c c c c }
 \toprule
          $ID$ & $RA$ & $Dec $ & $z$ & $\log L_\text{1350}$  & $\log L_\text{AGN}$  & $Scan~ ID$ \\
              & (deg) & (deg) & & (erg/s) & (erg/s) \\ 
         \midrule
SDSS J015925.07$-$001755.4 & 29.8545 & $-$0.2987 & 3.2570 & 46.72 & 47.02 & HerS\\
SDSS J020719.65$-$001959.8 & 31.8319 & $-$0.3333 & 3.4013 & 46.80 & 47.11 & HerS\\
SDSS J020948.58$+$002726.6 & 32.4524 & 0.4574 & 2.6929 & 46.69 & 46.99 & HerS\\
SDSS J020950.71$-$000506.4 & 32.4613 & $-$0.0851 & 2.8282 & 47.33 & 47.63 & HerS\\
SDSS J021724.53$-$010357.5 & 34.3522 & $-$1.0660 & 2.2345 & 46.51 & 46.81 & HerS\\
SDSS J022205.54$+$004335.2 & 35.5231 & 0.7265 & 2.5259 & 46.50 & 46.80 & HerS\\
HE 0251$-$5550 & 43.1672 & $-$55.6422 & 2.3505 & 47.27 & 47.57 & 1342270329\\
SDSS J031712.23$-$075850.3 & 49.3010 & $-$7.9807 & 2.6957 & 46.59 & 46.90 & 1342239839\\
SDSS J075547.83$+$220450.1 & 118.9493 & 22.0806 & 2.3210 & 46.92 & 47.22 & 1342270319\\
SDSS J081127.44$+$461812.9 & 122.8644 & 46.3036 & 2.2592 & 47.16 & 47.46 & 1342270275\\
SDSS J081940.58$+$082357.9 & 124.9191 & 8.3994 & 3.2147 & 46.67 & 46.97 & 1342270311\\
SDSS J082138.94$+$121729.9 & 125.4123 & 12.2917 & 3.1128 & 46.56 & 46.86 & 1342254515\\
SDSS J083249.39$+$155408.6 & 128.2058 & 15.9024 & 2.4165 & 46.55 & 46.85 & 1342270302\\
SDSS J084846.10$+$611234.6 & 132.1921 & 61.2096 & 2.2558 & 47.20 & 47.50 & 1342270242\\
SDSS J085417.61$+$532735.2 & 133.5734 & 53.4598 & 2.4182 & 46.92 & 47.23 & 1342270247\\
SDSS J085825.71$+$005006.7 & 134.6071 & 0.8352 & 2.8550 & 46.56 & 46.87 & H$-$ATLAS SDP\\
SDSS J085856.00$+$015219.4 & 134.7334 & 1.8721 & 2.1566 & 46.82 & 47.12 & H$-$ATLAS SDP\\
SDSS J085959.14$+$020519.7 & 134.9964 & 2.0888 & 2.9804 & 46.88 & 47.18 & H$-$ATLAS SDP\\
SDSS J090444.33$+$233354.0 & 136.1847 & 23.5650 & 2.2570 & 46.93 & 47.23 & 1342270297\\
SDSS J091054.79$+$023704.5 & 137.7283 & 2.6179 & 3.2951 & 46.75 & 47.05 & H$-$ATLAS SDP\\
SDSS J091247.59$-$004717.3 & 138.1983 & $-$0.7882 & 2.8593 & 46.55 & 46.86 & H$-$ATLAS SDP\\     
\end{tabular}}
\end{sidewaystable}

\newpage 
\addtocounter{table}{-1}
\begin{sidewaystable}
\caption{Continued.}
\vspace*{1cm}
\label{table_all_info3}
\tiny
\hspace{-2.5cm}
\resizebox{23.5cm}{!}{
\begin{tabular}{ l c c c c c c }
 \toprule
            $ID$ & $RA$ & $Dec $ & $z$ & $\log L_\text{1350}$  & $\log L_\text{AGN}$  & $Scan~ ID$ \\
             & (deg) & (deg) & & (erg/s) & (erg/s) \\ 
         \midrule
SDSS J092024.44$+$662656.7 & 140.1019 & 66.4491 & 2.0187 & 46.62 & 46.92 & 1342229122\\
SDSS J092325.25$+$453222.2 & 140.8552 & 45.5395 & 3.4524 & 47.07 & 47.37 & 1342270256\\
SDSS J092849.24$+$504930.5 & 142.2052 & 50.8252 & 2.3488 & 46.64 & 46.94 & 1342230874\\
SDSS J095112.84$+$025527.3 & 147.8035 & 2.9242 & 2.3732 & 46.52 & 46.83 & 1342209294\\
SDSS J095434.93$+$091519.6 & 148.6456 & 9.2554 & 3.3817 & 46.69 & 46.99 & 1342197310\\
SDSS J100515.99$+$480533.3 & 151.3166 & 48.0926 & 2.3850 & 47.01 & 47.31 & 1342270254\\
SDSS J101120.39$+$031244.5 & 152.8350 & 3.2124 & 2.4580 & 46.93 & 47.23 & 1342198868\\
SDSS J102325.31$+$514251.0 & 155.8555 & 51.7142 & 3.4510 & 47.18 & 47.48 & 1342270253\\
SDSS J102719.13$+$584114.3 & 156.8297 & 58.6873 & 2.0248 & 46.54 & 46.84 & 1342245910\\
SDSS J104018.51$+$572448.1 & 160.0772 & 57.4134 & 3.4089 & 46.88 & 47.18 & Lockman$-$North\\
SDSS J104121.88$+$563001.2 & 160.3412 & 56.5003 & 2.0519 & 46.61 & 46.91 & Lockman$-$Swire\\
SDSS J104442.15$+$381257.2 & 161.1756 & 38.2159 & 2.0745 & 46.57 & 46.87 & 1342254049\\
SDSS J104639.43$+$584047.7 & 161.6643 & 58.6799 & 3.1801 & 46.73 & 47.03 & Lockman$-$North\\
SDSS J104809.19$+$570241.9 & 162.0383 & 57.0450 & 3.2487 & 46.77 & 47.08 & Lockman$-$North\\
SDSS J105146.05$+$592214.0 & 162.9419 & 59.3706 & 2.9040 & 46.57 & 46.87 & Lockman$-$Swire\\
SDSS J105902.04$+$580848.6 & 164.7585 & 58.1469 & 2.2444 & 46.69 & 46.99 & Lockman$-$Swire\\
SDSS J110445.39$+$573643.9 & 166.1892 & 57.6122 & 2.6419 & 46.54 & 46.84 & Lockman$-$Swire\\
SDSS J111313.29$+$102212.4 & 168.3054 & 10.3701 & 2.2475 & 46.72 & 47.02 & 1342199324\\
SDSS J111928.37$+$130251.0 & 169.8682 & 13.0475 & 2.3940 & 46.68 & 46.98 & 1342198883\\
SDSS J113157.72$+$191527.7 & 172.9905 & 19.2577 & 2.9153 & 46.80 & 47.10 & 1342256846\\
SDSS J113627.81$+$541504.4 & 174.1159 & 54.2512 & 3.2360 & 46.61 & 46.92 & 1342195958\\    
\end{tabular}}
\end{sidewaystable}

\newpage

\addtocounter{table}{-1}
\begin{sidewaystable}
\caption{Continued.}
\vspace*{1cm}
\label{table_all_info4}
\tiny
\hspace{-2.5cm}
\resizebox{23.5cm}{!}{
\begin{tabular}{ l c c c c c c }
 \toprule
$ID$ & $RA$ & $Dec $ & $z$ & $\log L_\text{1350}$  & $\log L_\text{AGN}$  & $Scan~ ID$ \\
             & (deg) & (deg) & & (erg/s) & (erg/s) \\ 
         \midrule
SDSS J114412.76$+$315800.8 & 176.0532 & 31.9669 & 3.2350 & 46.76 & 47.06 & 1342256832\\
SDSS J115517.34$+$634622.0 & 178.8223 & 63.7728 & 2.8882 & 46.86 & 47.16 & 1342256631\\
SDSS J122307.52$+$103448.1 & 185.7813 & 10.5801 & 2.7422 & 46.58 & 46.89 & 1342234890\\
SDSS J122654.39$-$005430.6 & 186.7266 & $-$0.9085 & 2.6170 & 46.66 & 46.97 & 1342234883\\
SDSS J123132.37$+$013814.0 & 187.8849 & 1.6373 & 3.2286 & 46.77 & 47.08 & 1342257370\\
SDSS J123515.83$+$630113.3 & 188.8160 & 63.0204 & 2.3885 & 46.93 & 47.23 & 1342270217\\
SDSS J123637.45$+$615814.3 & 189.1560 & 61.9707 & 2.5199 & 46.55 & 46.85 & GOODS$-$North\\
SDSS J123714.60$+$064759.5 & 189.3108 & 6.7999 & 2.7811 & 46.54 & 46.84 & 1342234888\\
SDSS J123743.08$+$630144.8 & 189.4295 & 63.0291 & 3.4250 & 46.62 & 46.92 & 1342256809\\
SDSS J124302.42$+$521009.8 & 190.7601 & 52.1694 & 2.5588 & 46.51 & 46.81 & 1342198244\\
SDSS J124456.98$+$620143.0 & 191.2374 & 62.0286 & 3.0569 & 46.66 & 46.97 & 1342256811\\
SDSS J124748.44$+$042627.1 & 191.9519 & 4.4409 & 2.7833 & 46.58 & 46.88 & 1342189442\\
SDSS J125125.36$+$412000.4 & 192.8557 & 41.3335 & 3.1734 & 46.60 & 46.90 & 1342188754\\
SDSS J125819.24$+$165717.6 & 194.5802 & 16.9549 & 2.7015 & 46.64 & 46.94 & 1342259439\\
SDSS J131215.22$+$423900.8 & 198.0635 & 42.6502 & 2.5668 & 46.62 & 46.92 & 1342248486\\
SDSS J132809.59$+$545452.7 & 202.0400 & 54.9147 & 2.0958 & 46.81 & 47.12 & 1342256892\\
SDSS J133219.65$+$622715.9 & 203.0819 & 62.4544 & 3.1783 & 46.52 & 46.83 & 1342256897\\
SDSS J133907.13$+$131039.6 & 204.7797 & 13.1777 & 2.2411 & 46.65 & 46.95 & 1342259446\\
SDSS J135559.03$-$002413.6 & 208.9960 & $-$0.4038 & 2.3366 & 46.59 & 46.89 & 1342202220\\
SDSS J141819.22$+$044135.0 & 214.5801 & 4.6931 & 2.5006 & 46.79 & 47.09 & 1342213465\\
SDSS J142539.01$+$331009.5 & 216.4125 & 33.1693 & 2.3056 & 46.58 & 46.88 & Bo\"{o}tes\\   

\end{tabular}}
\end{sidewaystable}

\newpage 
\addtocounter{table}{-1}
\begin{sidewaystable}
\caption{Continued.}
\vspace*{1cm}
\label{table_all_info5}
\tiny
\hspace{-2.5cm}
\resizebox{23.5cm}{!}{
\begin{tabular}{ l c c c c c c }
 \toprule
            $ID$ & $RA$ & $Dec $ & $z$ & $\log L_\text{1350}$  & $\log L_\text{AGN}$  & $Scan~ ID$ \\
              & (deg) & (deg) & & (erg/s) & (erg/s) \\ 
         \midrule
SDSS J142539.98$+$344843.5 & 216.4166 & 34.8121 & 2.2516 & 46.54 & 46.84 & Bo\"{o}tes\\
SDSS J142912.87$+$340959.0 & 217.3037 & 34.1664 & 2.2289 & 46.66 & 46.97 & Bo\"{o}tes\\
SDSS J143543.71$+$342906.4 & 218.9322 & 34.4851 & 2.5731 & 46.68 & 46.98 & Bo\"{o}tes\\
SDSS J143941.92$+$332519.5 & 219.9247 & 33.4221 & 2.2536 & 46.54 & 46.84 & Bo\"{o}tes\\
SDSS J143954.64$+$334658.9 & 219.9777 & 33.7831 & 3.4390 & 46.63 & 46.93 & Bo\"{o}tes\\
SDSS J145706.34$+$220548.6 & 224.2764 & 22.0969 & 3.1114 & 46.52 & 46.82 & 1342201450\\
SDSS J155744.01$+$330231.0 & 239.4334 & 33.0420 & 3.1380 & 46.88 & 47.18 & 1342229549\\
SDSS J161238.26$+$532255.0 & 243.1594 & 53.3820 & 2.1392 & 46.76 & 47.07 & ELAIS\\
SDSS J210831.56$-$063022.5 & 317.1315 & $-$6.5063 & 2.3447 & 47.07 & 47.38 & 1342270337\\
SDSS J212329.46$-$005052.9 & 320.8728 & $-$0.8480 & 2.2614 & 47.37 & 47.67 & 1342270338\\
LBQS 2154$-$2005 & 329.2747 & $-$19.8538 & 2.0350 & 46.68 & 46.98 & 1342270203\\
HE 2156$-$4020 & 329.9779 & $-$40.0972 & 2.5310 & 47.02 & 47.32 & 1342270330\\
2QZ J221814.4$-$300306 & 334.5603 & $-$30.0517 & 2.3836 & 46.69 & 46.99 & 1342270331\\
2QZ J222006.7$-$280324 & 335.0279 & $-$28.0564 & 2.4060 & 47.37 & 47.67 & 1342270332\\
SDSS J222256.11$-$094636.2 & 335.7338 & $-$9.7767 & 2.9264 & 46.91 & 47.22 & 1342219976\\
SDSS J233446.40$-$090812.2 & 353.6933 & $-$9.1367 & 3.3169 & 47.09 & 47.39 & 1342234748\\     
\bottomrule
\end{tabular}}
\end{sidewaystable}

\newpage
\begin{sidewaystable}
\caption{The $z=2-3.5$ sample: $WISE$ and $Herschel$ fluxes and luminosities derived from SED fittings.}
\vspace*{1cm}
\label{table_detections_all_info}
\tiny
\hspace{-2.5cm}
\resizebox{23.5cm}{!}{
\begin{tabular}{ lccccccccc }
 \toprule
  $ID$& $F_\text{3.4$\mu$\rm m}$  & $F_\text{4.6$\mu$\rm m}$ & $F_\text{12$\mu$\rm m}$ & $F_\text{22$\mu$\rm m}$ & $F_\text{250$\mu$\rm m}$$^{\Delta}$ & $F_\text{350$\mu$\rm m}$$^{\Delta}$ & $F_\text{500$\mu$\rm m}$$^{\Delta}$& $\log L_\text{5$\mu$\rm m}$ & $\log L_\text{SF}$ \\
          & (mJy) & (mJy) & (mJy) & (mJy) & (mJy) & (mJy) & (mJy)  & (erg/s)  & $\Lsun$\\ 
\hline
SDSS J002025.22$+$154054.7 & 0.76$\pm$0.02 & 1.07$\pm$0.03 & 2.50$\pm$0.17 & 3.28$\pm$1.17 & 0.00 & 0.00 & 0.00 & 46.31$\pm$0.05 & $-$\\
SDSS J005202.40$+$010129.2 & 0.54$\pm$0.02 & 0.77$\pm$0.03 & 2.49$\pm$0.27 & 6.06$\pm$1.68 & 0.00 & 0.00 & 0.00 & 46.46$\pm$0.13 & $-$\\
SDSS J005229.51$-$110309.9 & 0.22$\pm$0.01 & 0.33$\pm$0.02 & 1.70$\pm$0.13 & 2.74$\pm$1.07 & 43.38$\pm$2.60 & 41.13$\pm$2.63 & 31.18$\pm$3.27 & 46.28$\pm$0.04 & 13.03$^{+0.01}_{-0.01}$\\
SDSS J005814.31$+$011530.2 & 0.35$\pm$0.01 & 0.52$\pm$0.02 & 2.63$\pm$0.20 & 6.96$\pm$1.20 & 0.00 & 0.00 & 0.00 & 46.60$\pm$0.16 & $-$\\
SDSS J010227.51$+$005136.8 & 0.18$\pm$0.01 & 0.30$\pm$0.02 & 1.41$\pm$0.17 & 3.96$\pm$1.40 & 0.00 & 0.00 & 0.00 & 46.35$\pm$0.18 & $-$\\
SDSS J010612.21$+$001920.1 & 0.19$\pm$0.01 & 0.21$\pm$0.01 & 1.16$\pm$0.17 & 4.12$\pm$1.11 & 0.00 & 0.00 & 0.00 & 46.54$\pm$0.26 & $-$\\
SDSS J011552.59$+$000601.0 & 0.09$\pm$0.01 & 0.11$\pm$0.01 & 0.63$\pm$0.27 & 3.10$\pm$1.70 & 0.00 & 0.00 & 0.00 & 46.37$\pm$0.39\textsuperscript{\ddag} & $-$\\
SDSS J011827.99$-$005239.8 & 0.31$\pm$0.01 & 0.53$\pm$0.02 & 2.42$\pm$0.14 & 5.82$\pm$1.14 & 37.86$\pm$6.60 & 22.50$\pm$6.15 & 23.12$\pm$8.38 & 46.40$\pm$0.13 & 12.88$^{+0.06}_{-0.07}$\\
SDSS J012412.46$-$010049.8 & 1.25$\pm$0.03 & 1.34$\pm$0.03 & 3.47$\pm$0.15 & 7.76$\pm$1.02 & 0.00 & 0.00 & 0.00 & 46.81$\pm$0.08 & $-$\\
SDSS J012517.14$-$001828.9 & 0.24$\pm$0.01 & 0.30$\pm$0.01 & 1.00$\pm$0.19 & 2.16$\pm$1.14 & 0.00 & 0.00 & 0.00 & 46.03$\pm$0.09 & $-$\\
SDSS J012748.31$-$001333.0 & 0.46$\pm$0.01 & 1.01$\pm$0.03 & 6.08$\pm$0.16 & 19.91$\pm$1.08 & 65.49$\pm$5.97 & 24.57$\pm$6.32 & 12.60$\pm$7.68 & 46.82$\pm$0.25\textsuperscript{\ddag} & 13.01$^{+0.04}_{-0.04}$\\
SDSS J013014.30$-$000639.2 & 0.23$\pm$0.01 & 0.37$\pm$0.02 & 1.92$\pm$0.13 & 2.42$\pm$0.92 & 0.00 & 0.00 & 0.00 & 46.25$\pm$0.13 & $-$\\
SDSS J013249.38$+$002627.1 & 0.14$\pm$0.01 & 0.17$\pm$0.01 & 0.53$\pm$0.15 & 2.92\textsuperscript{\dag}  & 0.00 & 0.00 & 0.00 & 46.07$\pm$0.18 & $-$\\
SDSS J013654.33$-$003415.4 & 0.13$\pm$0.01 & 0.16$\pm$0.01 & 0.46$\pm$0.13 & 3.59\textsuperscript{\dag}  & 35.16$\pm$7.56 & 19.27$\pm$7.97 & 18.70$\pm$9.31 & 45.85$\pm$0.17 & 13.01$^{+0.07}_{-0.09}$\\
SDSS J014123.04$-$002422.0 & 0.30$\pm$0.01 & 0.30$\pm$0.01 & 0.62$\pm$0.17 & 2.22$\pm$1.34 & 0.00 & 0.00 & 0.00 & 46.07$\pm$0.27 & $-$\\
SDSS J014214.75$+$002324.2 & 0.32$\pm$0.01 & 0.28$\pm$0.01 & 0.97$\pm$0.12 & 4.04\textsuperscript{\dag}  & 0.00 & 0.00 & 0.00 & 46.57$\pm$0.32 & $-$\\
SDSS J014303.16$+$001039.6 & 0.25$\pm$0.01 & 0.31$\pm$0.01 & 1.68$\pm$0.17 & 3.90$\pm$1.16 & 0.00 & 0.00 & 0.00 & 46.37$\pm$0.11 & $-$\\
SDSS J014733.58$+$000323.2 & 0.42$\pm$0.01 & 0.70$\pm$0.02 & 2.72$\pm$0.16 & 6.23$\pm$0.94 & 0.00 & 0.00 & 0.00 & 46.33$\pm$0.14 & $-$\\
SDSS J014809.64$-$001017.8 & 0.36$\pm$0.01 & 0.65$\pm$0.02 & 2.76$\pm$0.12 & 7.00$\pm$0.84 & 81.65$\pm$7.47 & 73.49$\pm$7.56 & 76.10$\pm$9.12 & 46.45$\pm$0.15 & 13.23$^{+0.02}_{-0.02}$\\
SDSS J015017.71$+$002902.4 & 0.12$\pm$0.01 & 0.20$\pm$0.01 & 1.44$\pm$0.11 & 2.83$\pm$1.01 & 66.22$\pm$6.21 & 83.78$\pm$6.43 & 60.77$\pm$7.58 & 46.45$\pm$0.03 & 13.43$^{+0.02}_{-0.02}$\\
SDSS J015819.77$-$001222.0 & 0.18$\pm$0.01 & 0.21$\pm$0.01 & 1.36$\pm$0.11 & 4.37$\pm$0.87 & 40.66$\pm$7.70 & 25.93$\pm$8.01 & 9.13$\pm$9.22 & 46.65$\pm$0.22 & 13.08$^{+0.08}_{-0.09}$\\
\cmidrule{2-10}
\end{tabular}}
 \begin{tablenotes}
 \item \dag $2\,\sigma$ upper confidence  limit in the corresponding $WISE$ band.
 \item \ddag  Weak near-infra-red source.
   \item $^{\Delta}$This value is zero if the source was not detected at a significance larger than $3\,\sigma$ at 250$\mu \rm m$.
 \end{tablenotes}       
\end{sidewaystable}

\newpage
\addtocounter{table}{-1}
\begin{sidewaystable}
\caption{Continued.}
\vspace*{1cm}
\label{table_detections_all_info2}
\tiny
\hspace{-2.5cm}
\resizebox{23.5cm}{!}{
\begin{tabular}{ lccccccccc }
 \toprule
 $ID$& $F_\text{3.4$\mu$\rm m}$  & $F_\text{4.6$\mu$\rm m}$ & $F_\text{12$\mu$\rm m}$ & $F_\text{22$\mu$\rm m}$ & $F_\text{250$\mu$\rm m}$$^{\Delta}$ & $F_\text{350$\mu$\rm m}$$^{\Delta}$ & $F_\text{500$\mu$\rm m}$$^{\Delta}$& $\log L_\text{5$\mu$\rm m}$ & $\log L_\text{SF}$ \\
          & (mJy) & (mJy) & (mJy) & (mJy) & (mJy) & (mJy) & (mJy)  & (erg/s)  & $\Lsun$\\ 
\hline
SDSS J015925.07$-$001755.4 & 0.14$\pm$0.01 & 0.16$\pm$0.01 & 1.05$\pm$0.12 & 2.07$\pm$0.90 & 0.00 & 0.00 & 0.00 & 46.41$\pm$0.03 & $-$\\
SDSS J020719.65$-$001959.8 & 0.25$\pm$0.01 & 0.22$\pm$0.01 & 0.44$\pm$0.12 & 3.18\textsuperscript{\dag}  & 0.00 & 0.00 & 0.00 & 46.36$\pm$0.53 & $-$\\
SDSS J020948.58$+$002726.6 & 0.21$\pm$0.01 & 0.29$\pm$0.01 & 1.87$\pm$0.12 & 6.05$\pm$1.06 & 0.00 & 0.00 & 0.00 & 46.57$\pm$0.23 & $-$\\
SDSS J020950.71$-$000506.4 & 0.92$\pm$0.02 & 1.33$\pm$0.03 & 5.63$\pm$0.17 & 15.40$\pm$0.95 & 75.47$\pm$10.91 & 64.20$\pm$10.43 & 41.81$\pm$11.48 & 47.07$\pm$0.16 & 13.24$^{+0.11}_{-0.15}$\\
SDSS J021724.53$-$010357.5 & 0.15$\pm$0.01 & 0.26$\pm$0.01 & 1.03$\pm$0.14 & 2.65$\pm$1.13 & 0.00 & 0.00 & 0.00 & 46.06$\pm$0.16 & $-$\\
SDSS J022205.54$+$004335.2 & 0.12$\pm$0.01 & 0.17$\pm$0.01 & 1.04$\pm$0.13 & 2.11$\pm$1.03 & 0.00 & 0.00 & 0.00 & 46.14$\pm$0.05 & $-$\\
HE 0251$-$5550 & 1.70$\pm$0.04 & 1.91$\pm$0.04 & 7.57$\pm$0.18 & 18.63$\pm$1.00 & 30.76$\pm$6.37 & 27.70$\pm$6.90 & 15.62$\pm$7.57 & 46.98$\pm$0.13 & 12.41$^{+0.23}_{-0.18}$\\
SDSS J031712.23$-$075850.3 & 0.12$\pm$0.01 & 0.13$\pm$0.01 & 0.46$\pm$0.17 & 2.05$\pm$1.29 & 0.00 & 0.00 & 0.00 & 46.03$\pm$0.36 & $-$\\
SDSS J075547.83$+$220450.1 & 0.58$\pm$0.02 & 0.85$\pm$0.03 & 2.98$\pm$0.18 & 5.70$\pm$1.04 & 0.00 & 0.00 & 0.00 & 46.52$\pm$0.02 & $-$\\
SDSS J081127.44$+$461812.9 & 0.51$\pm$0.01 & 0.89$\pm$0.02 & 3.26$\pm$0.15 & 7.13$\pm$1.08 & 44.00$\pm$6.40 & 25.99$\pm$6.87 & 0.00 & 46.55$\pm$0.09 & 12.92$^{+0.06}_{-0.09}$\\
SDSS J081940.58$+$082357.9 & 0.67$\pm$0.02 & 0.72$\pm$0.02 & 2.49$\pm$0.20 & 6.88$\pm$1.89 & 0.00 & 0.00 & 0.00 & 46.85$\pm$0.16 & $-$\\
SDSS J082138.94$+$121729.9 & 0.62$\pm$0.02 & 0.52$\pm$0.02 & 0.81$\pm$0.18 & 3.05$\pm$1.62 & 0.00 & 0.00 & 0.00 & 46.39$\pm$0.28 & $-$\\
SDSS J083249.39$+$155408.6 & 0.34$\pm$0.01 & 0.55$\pm$0.03 & 4.17$\pm$0.24 & 7.22$\pm$1.83 & 0.00 & 0.00 & 0.00 & 46.67$\pm$0.02 & $-$\\
SDSS J084846.10$+$611234.6 & 1.28$\pm$0.03 & 1.49$\pm$0.03 & 3.79$\pm$0.15 & 8.78$\pm$0.95 & 0.00 & 0.00 & 0.00 & 46.63$\pm$0.11 & $-$\\
SDSS J085417.61$+$532735.2 & 1.01$\pm$0.02 & 1.23$\pm$0.03 & 3.70$\pm$0.17 & 7.27$\pm$1.04 & 0.00 & 0.00 & 0.00 & 46.65$\pm$0.04 & $-$\\
SDSS J085825.71$+$005006.7 & 0.13$\pm$0.01 & 0.18$\pm$0.01 & 0.82$\pm$0.16 & 2.35$\pm$1.45 & 0.00 & 0.00 & 0.00 & 46.25$\pm$0.18 & $-$\\
SDSS J085856.00$+$015219.4 & 0.46$\pm$0.01 & 0.80$\pm$0.02 & 3.48$\pm$0.16 & 8.63$\pm$1.03 & 0.00 & 0.00 & 0.00 & 46.56$\pm$0.14 & $-$\\
SDSS J085959.14$+$020519.7 & 0.15$\pm$0.01 & 0.22$\pm$0.01 & 1.02$\pm$0.13 & 3.27$\pm$1.06 & 0.00 & 0.00 & 0.00 & 46.41$\pm$0.22 & $-$\\
SDSS J090444.33$+$233354.0 & 1.07$\pm$0.03 & 1.39$\pm$0.03 & 4.57$\pm$0.17 & 8.57$\pm$1.24 & 0.00 & 0.00 & 0.00 & 46.66$\pm$0.03 & $-$\\
SDSS J091054.79$+$023704.5 & 0.15$\pm$0.01 & 0.14$\pm$0.01 & 0.38$\pm$0.16 & 2.88$\pm$1.15 & 0.00 & 0.00 & 0.00 & 46.27$\pm$0.56\textsuperscript{\ddag} & $-$\\
SDSS J091247.59$-$004717.3 & 0.14$\pm$0.01 & 0.22$\pm$0.01 & 1.21$\pm$0.17 & 2.76$\pm$1.25 & 0.00 & 0.00 & 0.00 & 46.37$\pm$0.09 & $-$\\

\cmidrule{2-10}
\end{tabular}}
 \begin{tablenotes}
 \item \dag $2\,\sigma$ upper confidence  limit in the corresponding $WISE$ band.
 \item \ddag  Weak near-infra-red source.
   \item $^{\Delta}$This value is zero if the source was not detected at a significance larger than $3\,\sigma$ at 250$\mu \rm m$.
 \end{tablenotes}       
\end{sidewaystable}

\newpage
\addtocounter{table}{-1}
\begin{sidewaystable}
\caption{Continued.}
\vspace*{1cm}
\label{table_detections_all_info3}
\tiny
\hspace{-2.5cm}
\resizebox{23.5cm}{!}{
\begin{tabular}{ lccccccccc }
 \toprule
 $ID$& $F_\text{3.4$\mu$\rm m}$  & $F_\text{4.6$\mu$\rm m}$ & $F_\text{12$\mu$\rm m}$ & $F_\text{22$\mu$\rm m}$ & $F_\text{250$\mu$\rm m}$$^{\Delta}$ & $F_\text{350$\mu$\rm m}$$^{\Delta}$ & $F_\text{500$\mu$\rm m}$$^{\Delta}$& $\log L_\text{5$\mu$\rm m}$ & $\log L_\text{SF}$ \\
          & (mJy) & (mJy) & (mJy) & (mJy) & (mJy) & (mJy) & (mJy)  & (erg/s)  & $\Lsun$\\ 
\hline
SDSS J092024.44$+$662656.7 & 0.45$\pm$0.01 & 0.87$\pm$0.03 & 3.33$\pm$0.13 & 6.14$\pm$0.93 & 80.71$\pm$4.47 & 70.74$\pm$4.99 & 29.39$\pm$5.62 & 46.38$\pm$0.07 & 13.14$^{+0.01}_{-0.02}$\\
SDSS J092325.25$+$453222.2 & 0.47$\pm$0.01 & 0.39$\pm$0.02 & 1.44$\pm$0.16 & 4.35$\pm$1.50 & 0.00 & 0.00 & 0.00 & 46.70$\pm$0.19 & $-$\\
SDSS J092849.24$+$504930.5 & 0.32$\pm$0.01 & 0.50$\pm$0.02 & 2.37$\pm$0.12 & 5.79$\pm$0.82 & 75.14$\pm$4.23 & 60.68$\pm$4.49 & 44.55$\pm$5.26 & 46.47$\pm$0.13 & 13.21$^{+0.01}_{-0.02}$\\
SDSS J095112.84$+$025527.3 & 0.19$\pm$0.01 & 0.28$\pm$0.01 & 1.03$\pm$0.16 & 4.15\textsuperscript{\dag}  & 0.00 & 0.00 & 0.00 & 46.05$\pm$0.18 & $-$\\
SDSS J095434.93$+$091519.6 & 0.14$\pm$0.01 & 0.16$\pm$0.01 & 1.07$\pm$0.18 & 2.53$\pm$1.26 & 78.44$\pm$4.09 & 101.53$\pm$3.56 & 88.02$\pm$4.04 & 46.50$\pm$0.10 & 13.66$^{+0.01}_{-0.01}$\\
SDSS J100515.99$+$480533.3 & 0.38$\pm$0.01 & 0.46$\pm$0.02 & 1.62$\pm$0.12 & 3.56$\pm$0.99 & 0.00 & 0.00 & 0.00 & 46.30$\pm$0.09 & $-$\\
SDSS J101120.39$+$031244.5 & 0.58$\pm$0.02 & 0.74$\pm$0.02 & 2.60$\pm$0.17 & 5.28$\pm$1.20 & 0.00 & 0.00 & 0.00 & 46.52$\pm$0.05 & $-$\\
SDSS J102325.31$+$514251.0 & 0.43$\pm$0.01 & 0.39$\pm$0.01 & 1.42$\pm$0.12 & 3.51$\pm$1.10 & 20.96$\pm$6.39 & 18.39$\pm$6.89 & 22.52$\pm$7.44 & 46.65$\pm$0.11 & 12.84$^{+0.13}_{-0.16}$\\
SDSS J102719.13$+$584114.3 & 0.24$\pm$0.01 & 0.51$\pm$0.02 & 2.10$\pm$0.12 & 3.15$\pm$1.09 & 0.00 & 0.00 & 0.00 & 46.14$\pm$0.07 & $-$\\
SDSS J104018.51$+$572448.1 & 0.41$\pm$0.01 & 0.38$\pm$0.01 & 1.03$\pm$0.11 & 1.87$\pm$0.85 & 0.00 & 0.00 & 0.00 & 46.47$\pm$0.03 & $-$\\
SDSS J104121.88$+$563001.2 & 0.25$\pm$0.01 & 0.45$\pm$0.01 & 2.21$\pm$0.14 & 7.63$\pm$0.97 & 0.00 & 0.00 & 0.00 & 46.27$\pm$0.36\textsuperscript{\ddag} & $-$\\
SDSS J104442.15$+$381257.2 & 0.20$\pm$0.01 & 0.27$\pm$0.01 & 0.79$\pm$0.15 & 3.97\textsuperscript{\dag}  & 0.00 & 0.00 & 0.00 & 45.79$\pm$0.18 & $-$\\
SDSS J104639.43$+$584047.7 & 0.13$\pm$0.01 & 0.16$\pm$0.01 & 1.08$\pm$0.12 & 2.92$\pm$0.91 & 0.00 & 0.00 & 0.00 & 46.40$\pm$0.07 & $-$\\
SDSS J104809.19$+$570241.9 & 0.18$\pm$0.01 & 0.24$\pm$0.01 & 1.60$\pm$0.12 & 5.01$\pm$0.91 & 0.00 & 0.00 & 0.00 & 46.63$\pm$0.07\textsuperscript{\ddag} & $-$\\
SDSS J105146.05$+$592214.0 & 0.17$\pm$0.01 & 0.21$\pm$0.01 & 0.69$\pm$0.14 & 2.62$\pm$1.25 & 0.00 & 0.00 & 0.00 & 46.25$\pm$0.29 & $-$\\
SDSS J105902.04$+$580848.6 & 0.45$\pm$0.01 & 0.64$\pm$0.02 & 2.39$\pm$0.14 & 5.69$\pm$1.13 & 24.39$\pm$2.06 & 19.06$\pm$1.98 & 0.00 & 46.42$\pm$0.13 & 12.69$^{+0.03}_{-0.03}$\\
SDSS J110445.39$+$573643.9 & 0.13$\pm$0.01 & 0.17$\pm$0.01 & 0.80$\pm$0.14 & 2.28$\pm$1.21 & 0.00 & 0.00 & 0.00 & 46.15$\pm$0.19 & $-$\\
SDSS J111313.29$+$102212.4 & 0.77$\pm$0.02 & 1.14$\pm$0.03 & 4.61$\pm$0.19 & 11.26$\pm$1.20 & 0.00 & 0.00 & 0.00 & 46.72$\pm$0.13 & $-$\\
SDSS J111928.37$+$130251.0 & 0.25$\pm$0.01 & 0.26$\pm$0.02 & 1.23$\pm$0.17 & 2.21$\pm$1.27 & 0.00 & 0.00 & 0.00 & 46.13$\pm$0.02 & $-$\\
SDSS J113157.72$+$191527.7 & 0.47$\pm$0.01 & 0.53$\pm$0.02 & 1.32$\pm$0.17 & 3.55$\pm$1.27 & 0.00 & 0.00 & 0.00 & 46.46$\pm$0.15 & $-$\\
SDSS J113627.81$+$541504.4 & 0.20$\pm$0.01 & 0.25$\pm$0.01 & 0.99$\pm$0.12 & 1.72$\pm$1.09 & 0.00 & 0.00 & 0.00 & 46.35$\pm$0.03 & $-$\\
\cmidrule{2-10}
\end{tabular}}
 \begin{tablenotes}
 \item \dag $2\,\sigma$ upper confidence  limit in the corresponding $WISE$ band.
 \item \ddag  Weak near-infra-red source.
   \item $^{\Delta}$This value is zero if the source was not detected at a significance larger than $3\,\sigma$ at 250$\mu \rm m$.
 \end{tablenotes}       
\end{sidewaystable}

\newpage
\addtocounter{table}{-1}
\begin{sidewaystable}
\caption{Continued.}
\vspace*{1cm}
\label{table_detections_all_info4}
\tiny
\hspace{-2.5cm}
\resizebox{23.5cm}{!}{
\begin{tabular}{ lccccccccc }
 \toprule
 $ID$& $F_\text{3.4$\mu$\rm m}$  & $F_\text{4.6$\mu$\rm m}$ & $F_\text{12$\mu$\rm m}$ & $F_\text{22$\mu$\rm m}$ & $F_\text{250$\mu$\rm m}$$^{\Delta}$ & $F_\text{350$\mu$\rm m}$$^{\Delta}$ & $F_\text{500$\mu$\rm m}$$^{\Delta}$& $\log L_\text{5$\mu$\rm m}$ & $\log L_\text{SF}$ \\
          & (mJy) & (mJy) & (mJy) & (mJy) & (mJy) & (mJy) & (mJy)  & (erg/s)  & $\Lsun$\\ 
\hline
SDSS J114412.76$+$315800.8 & 0.15$\pm$0.01 & 0.18$\pm$0.01 & 1.04$\pm$0.13 & 2.28$\pm$0.88 & 30.52$\pm$6.34 & 32.53$\pm$6.75 & 26.82$\pm$7.44 & 46.42$\pm$0.07 & 13.08$^{+0.06}_{-0.13}$\\
SDSS J115517.34$+$634622.0 & 0.90$\pm$0.02 & 0.77$\pm$0.02 & 1.54$\pm$0.12 & 4.05$\pm$1.01 & 33.40$\pm$7.82 & 16.84$\pm$8.18 & 13.66$\pm$9.55 & 46.52$\pm$0.15 & 12.90$^{+0.10}_{-0.14}$\\
SDSS J122307.52$+$103448.1 & 0.16$\pm$0.01 & 0.19$\pm$0.01 & 0.78$\pm$0.14 & 2.52\textsuperscript{\dag}  & 0.00 & 0.00 & 0.00 & 46.08$\pm$0.18 & $-$\\
SDSS J122654.39$-$005430.6 & 0.39$\pm$0.01 & 0.47$\pm$0.02 & 1.11$\pm$0.16 & 3.99$\pm$1.71 & 0.00 & 0.00 & 0.00 & 46.33$\pm$0.27 & $-$\\
SDSS J123132.37$+$013814.0 & 0.11$\pm$0.01 & 0.10$\pm$0.01 & 0.40$\pm$0.13 & 3.49$\pm$1.16 & 0.00 & 0.00 & 0.00 & 46.31$\pm$0.61\textsuperscript{\ddag} & $-$\\
SDSS J123515.83$+$630113.3 & 0.45$\pm$0.01 & 0.73$\pm$0.02 & 3.09$\pm$0.14 & 6.01$\pm$0.80 & 30.82$\pm$6.34 & 37.29$\pm$6.87 & 26.22$\pm$7.49 & 46.56$\pm$0.04 & 12.73$^{+0.06}_{-0.07}$\\
SDSS J123637.45$+$615814.3 & 0.17$\pm$0.01 & 0.24$\pm$0.01 & 0.80$\pm$0.11 & 1.87$\pm$0.99 & 0.00 & 0.00 & 0.00 & 46.06$\pm$0.11 & $-$\\
SDSS J123714.60$+$064759.5 & 0.43$\pm$0.01 & 0.63$\pm$0.02 & 3.80$\pm$0.15 & 11.81$\pm$0.99 & 94.20$\pm$3.68 & 94.26$\pm$3.88 & 54.41$\pm$4.56 & 46.91$\pm$0.21\textsuperscript{\ddag} & 13.43$^{+0.01}_{-0.00}$\\
SDSS J123743.08$+$630144.8 & 0.12$\pm$0.01 & 0.12$\pm$0.01 & 0.80$\pm$0.08 & 2.72$\pm$0.70 & 25.83$\pm$2.52 & 20.00$\pm$2.47 & 13.63$\pm$3.00 & 46.46$\pm$0.24 & 12.98$^{+0.03}_{-0.04}$\\
SDSS J124302.42$+$521009.8 & 0.16$\pm$0.01 & 0.23$\pm$0.01 & 1.14$\pm$0.13 & 2.53$\pm$1.31 & 0.00 & 0.00 & 0.00 & 46.22$\pm$0.09 & $-$\\
SDSS J124456.98$+$620143.0 & 0.16$\pm$0.01 & 0.17$\pm$0.01 & 0.90$\pm$0.09 & 2.00$\pm$0.80 & 0.00 & 0.00 & 0.00 & 46.30$\pm$0.08 & $-$\\
SDSS J124748.44$+$042627.1 & 0.16$\pm$0.01 & 0.18$\pm$0.01 & 0.37$\pm$0.18 & 2.65$\pm$1.33 & 0.00 & 0.00 & 0.00 & 46.07$\pm$0.55 & $-$\\
SDSS J125125.36$+$412000.4 & 0.12$\pm$0.01 & 0.16$\pm$0.01 & 1.34$\pm$0.12 & 3.30\textsuperscript{\dag}  & 48.71$\pm$3.58 & 39.74$\pm$3.90 & 27.13$\pm$4.85 & 46.48$\pm$0.17\textsuperscript{\ddag} & 13.25$^{+0.02}_{-0.02}$\\
SDSS J125819.24$+$165717.6 & 0.34$\pm$0.01 & 0.47$\pm$0.02 & 2.24$\pm$0.14 & 3.71$\pm$1.05 & 49.46$\pm$5.30 & 40.48$\pm$6.00 & 32.27$\pm$6.70 & 46.51$\pm$0.03 & 13.12$^{+0.03}_{-0.04}$\\
SDSS J131215.22$+$423900.8 & 0.23$\pm$0.01 & 0.29$\pm$0.01 & 1.22$\pm$0.13 & 3.18$\pm$1.00 & 0.00 & 0.00 & 0.00 & 46.29$\pm$0.15 & $-$\\
SDSS J132809.59$+$545452.7 & 0.26$\pm$0.01 & 0.44$\pm$0.01 & 2.12$\pm$0.11 & 5.73$\pm$0.78 & 77.97$\pm$6.34 & 73.04$\pm$6.79 & 46.97$\pm$7.48 & 46.33$\pm$0.18\textsuperscript{\ddag} & 13.09$^{+0.08}_{-0.03}$\\
SDSS J133219.65$+$622715.9 & 0.09$\pm$0.00 & 0.09$\pm$0.01 & 0.33$\pm$0.08 & 1.46$\pm$0.71 & 29.83$\pm$2.46 & 26.35$\pm$2.45 & 17.72$\pm$2.86 & 46.06$\pm$0.34\textsuperscript{\ddag} & 13.07$^{+0.02}_{-0.02}$\\
SDSS J133907.13$+$131039.6 & 0.42$\pm$0.01 & 0.48$\pm$0.02 & 1.05$\pm$0.13 & 3.47\textsuperscript{\dag}  & 0.00 & 0.00 & 0.00 & 46.00$\pm$0.18 & $-$\\
SDSS J135559.03$-$002413.6 & 0.26$\pm$0.01 & 0.41$\pm$0.02 & 1.63$\pm$0.10 & 2.89$\pm$0.82 & 19.03$\pm$2.66 & 24.26$\pm$2.66 & 0.00 & 46.23$\pm$0.02 & 12.67$^{+0.03}_{-0.04}$\\
SDSS J141819.22$+$044135.0 & 0.19$\pm$0.01 & 0.34$\pm$0.01 & 1.50$\pm$0.09 & 3.16$\pm$0.63 & 49.11$\pm$6.34 & 57.63$\pm$6.88 & 52.43$\pm$7.52 & 46.31$\pm$0.07 & 13.04$^{+0.04}_{-0.04}$\\
SDSS J142539.01$+$331009.5 & 0.26$\pm$0.01 & 0.28$\pm$0.01 & 0.55$\pm$0.11 & 2.49\textsuperscript{\dag}  & 0.00 & 0.00 & 0.00 & 45.75$\pm$0.18 & $-$\\
\cmidrule{2-10}
\end{tabular}}
 \begin{tablenotes}
 \item \dag $2\,\sigma$ upper confidence  limit in the corresponding $WISE$ band.
 \item \ddag  Weak near-infra-red source.
   \item $^{\Delta}$This value is zero if the source was not detected at a significance larger than $3\,\sigma$ at 250$\mu \rm m$.
 \end{tablenotes}       
\end{sidewaystable}

\newpage
\addtocounter{table}{-1}
\begin{sidewaystable}
\caption{Continued.}
\vspace*{1cm}
\label{table_detections_all_info5}
\tiny
\hspace{-2.5cm}
\resizebox{23.5cm}{!}{
\begin{tabular}{ lccccccccc }
 \toprule
 $ID$& $F_\text{3.4$\mu$\rm m}$  & $F_\text{4.6$\mu$\rm m}$ & $F_\text{12$\mu$\rm m}$ & $F_\text{22$\mu$\rm m}$ & $F_\text{250$\mu$\rm m}$$^{\Delta}$ & $F_\text{350$\mu$\rm m}$$^{\Delta}$ & $F_\text{500$\mu$\rm m}$$^{\Delta}$& $\log L_\text{5$\mu$\rm m}$ & $\log L_\text{SF}$ \\
          & (mJy) & (mJy) & (mJy) & (mJy) & (mJy) & (mJy) & (mJy)  & (erg/s)  & $\Lsun$\\ 
\hline
 SDSS J142539.98$+$344843.5 & 0.50$\pm$0.01 & 0.80$\pm$0.02 & 3.58$\pm$0.15 & 8.69$\pm$1.20 & 34.52$\pm$2.45 & 19.27$\pm$2.75 & 8.75$\pm$3.13 & 46.60$\pm$0.13\textsuperscript{\ddag} & 12.79$^{+0.03}_{-0.03}$\\
SDSS J142912.87$+$340959.0 & 0.24$\pm$0.01 & 0.37$\pm$0.01 & 1.51$\pm$0.09 & 5.19$\pm$0.73 & 34.42$\pm$1.29 & 25.05$\pm$1.23 & 15.48$\pm$1.51 & 46.29$\pm$0.27\textsuperscript{\ddag} & 12.88$^{+0.01}_{-0.01}$\\
SDSS J143543.71$+$342906.4 & 0.30$\pm$0.01 & 0.31$\pm$0.01 & 1.13$\pm$0.10 & 1.92$\pm$0.76 & 0.00 & 0.00 & 0.00 & 46.16$\pm$0.02 & $-$\\
SDSS J143941.92$+$332519.5 & 0.20$\pm$0.01 & 0.36$\pm$0.01 & 1.83$\pm$0.09 & 3.44$\pm$0.75 & 0.00 & 0.00 & 0.00 & 46.25$\pm$0.03 & $-$\\
SDSS J143954.64$+$334658.9 & 0.09$\pm$0.00 & 0.05$\pm$0.01 & 0.22$\pm$0.12 & 1.57$\pm$0.76 & 0.00 & 0.00 & 0.00 & 46.07$\pm$0.53 & $-$\\
SDSS J145706.34$+$220548.6 & 0.07$\pm$0.00 & 0.08$\pm$0.01 & 0.25$\pm$0.11 & 1.55\textsuperscript{\dag}  & 0.00 & 0.00 & 0.00 & 45.72$\pm$0.18 & $-$\\
SDSS J155744.01$+$330231.0 & 0.12$\pm$0.00 & 0.15$\pm$0.01 & 0.68$\pm$0.07 & 1.61$\pm$0.58 & 0.00 & 0.00 & 0.00 & 46.23$\pm$0.10 & $-$\\
SDSS J161238.26$+$532255.0 & 0.32$\pm$0.01 & 0.47$\pm$0.01 & 1.78$\pm$0.07 & 3.99$\pm$0.48 & 46.19$\pm$6.06 & 51.48$\pm$4.04 & 36.51$\pm$4.90 & 46.22$\pm$0.02 & 12.83$^{+0.03}_{-0.03}$\\
SDSS J210831.56$-$063022.5 & 0.49$\pm$0.01 & 0.82$\pm$0.02 & 4.10$\pm$0.19 & 7.72$\pm$1.06 & 41.10$\pm$6.36 & 27.65$\pm$6.90 & 20.45$\pm$7.54 & 46.65$\pm$0.03 & 12.82$^{+0.14}_{-0.16}$\\
SDSS J212329.46$-$005052.9 & 1.44$\pm$0.03 & 2.29$\pm$0.05 & 8.96$\pm$0.25 & 21.02$\pm$1.15 & 34.58$\pm$6.38 & 30.98$\pm$6.86 & 18.13$\pm$7.54 & 47.00$\pm$0.12 & 12.45$^{+0.19}_{-0.15}$\\
LBQS 2154$-$2005 & 0.51$\pm$0.01 & 0.92$\pm$0.03 & 4.34$\pm$0.20 & 8.93$\pm$1.33 & 34.67$\pm$6.38 & 17.30$\pm$6.85 & 0.00 & 46.50$\pm$0.06 & 12.73$^{+0.09}_{-0.24}$\\
HE 2156$-$4020 & 0.59$\pm$0.02 & 1.03$\pm$0.03 & 4.38$\pm$0.16 & 9.37$\pm$0.96 & 0.00 & 0.00 & 0.00 & 46.79$\pm$0.07 & $-$\\
2QZ J221814.4$-$300306 & 0.35$\pm$0.01 & 0.46$\pm$0.02 & 1.90$\pm$0.16 & 3.95$\pm$1.13 & 0.00 & 0.00 & 0.00 & 46.36$\pm$0.07 & $-$\\
2QZ J222006.7$-$280324 & 1.97$\pm$0.04 & 2.12$\pm$0.05 & 7.84$\pm$0.24 & 19.31$\pm$1.41 & 39.65$\pm$6.39 & 27.99$\pm$6.87 & 19.43$\pm$7.51 & 47.02$\pm$0.13 & 12.49$^{+0.26}_{-0.15}$\\
SDSS J222256.11$-$094636.2 & 0.42$\pm$0.01 & 0.44$\pm$0.02 & 1.85$\pm$0.15 & 4.53$\pm$1.33 & 0.00 & 0.00 & 0.00 & 46.59$\pm$0.12 & $-$\\
SDSS J233446.40$-$090812.2 & 0.29$\pm$0.01 & 0.31$\pm$0.01 & 1.43$\pm$0.16 & 2.21$\pm$1.05 & 50.77$\pm$6.90 & 56.18$\pm$7.36 & 37.35$\pm$8.30 & 46.51$\pm$0.07 & 13.35$^{+0.04}_{-0.06}$\\
\bottomrule
\end{tabular}}
 \begin{tablenotes}
 \item \dag $2\,\sigma$ upper confidence  limit in the corresponding $WISE$ band.
 \item \ddag  Weak near-infra-red source.
   \item $^{\Delta}$This value is zero if the source was not detected at a significance larger than $3\,\sigma$ at 250$\mu \rm m$.
 \end{tablenotes}       
\end{sidewaystable}


\begin{sidewaystable}
\caption{Fluxes and luminosities for median and mean stacks.}
\vspace*{1cm}
\label{table_stacks_all_info}
\tiny
\hspace{-1.5cm}
\resizebox{22.cm}{!}{
\begin{tabular}{ l c c c c c c c c c }
 \toprule
       &  $\log L_\text{1350}$ & $ \# \,\, of\,\, sources\,\, in\,\, the \,\,stack$ & $z_\text{nominal, stack} $ & $F_\text{250$\mu$\rm m, \rm stack}$ & $F_\text{350$\mu$\rm m, \rm stack}$ & $F_\text{500$\mu$\rm m, \rm stack}$ & $\log \lambda L_{\rm 250}$ & $\log \lambda L_{\rm 350}$ & $\log \lambda L_{\rm 500}$\\
        & (erg/s) & & & (mJy) & (mJy) & (mJy) & (erg/s) & (erg/s) & (erg/s) \\
         \midrule
     \bf{median} stack & $46.7-48$ & 28 & 2.5 & 8.16$[6.50, 9.95]$ & 6.15$[4.07, 7.68]$ & 3.89$[1.98, 5.95]$ & 45.13$[44.51, 45.38]$ & 45.18$[44.87, 45.36]$ & 44.45$[44.94, 45.17]$ \\
     \rule{0pt}{5ex}%
     & $46.5-46.7$ & 38 &   2.5 & 8.0$[6.27, 9.65]$ & 4.8$[3.52, 6.33]$ & 4.59$[3.26, 6.19]$  & 45.47$[45.29, 45.60]$ &45.17$[44.94, 45.32]$ & 45.09$[44.54, 45.32]$\\
          \hline
          \rule{0pt}{3.5ex}%
           \bf{mean} stack & $46.7-48$ & 28 & 2.5 & 8.16$[6.50, 9.95]$ & 6.15$[4.07, 7.68]$ & 3.89$[1.98, 5.95]$ & 45.62$[45.44, 45.74] $ & 45.24$[45.04, 45.38]$ & 45.17$[44.94, 45.31]$  \\
           \rule{0pt}{5ex}%
      & $46.5-46.7$ & 38 &   2.5 & 10.32$[7.88, 12.71]$ & 6.27$[4.76, 7.54]$ & 5.16$[3.64, 6.56]$ & 45.64$[45.46, 45.77]$ & 45.32$[45.18, 45.43]$ & 45.14$[44.98, 45.26]$ \\
\hline
\end{tabular}}
\end{sidewaystable}

\newpage

\begin{table}
\caption{Mean and median SFRs for detected and undetected \herschel\ sources and the entire sample. 
N is the number of sources in each group and the uncertainties reflect the 16 and 84 percentiles.}
\begin{tabular}{lrrrrrc} \hline
& \multicolumn{2}{c}{SPIRE Detected} && \multicolumn{2}{c}{SPIRE Stacks} & All sources\\
& mean & median && mean & median & mean\\ 
&&&&&\\
$\log L_{1350}<46.7$ & \multicolumn{2}{c}{N=19} && \multicolumn{2}{c}{N=38}& N=57\\
SFR ($M_{\odot}$/yr) & 1415$^{+707}_{-472}$ & 1160$^{+596}_{-485}$ && 151$^{+29}_{-33}$ & $115^{+26}_{-28}$ & 572\\
$\log L_{1350}>46.7$ & \multicolumn{2}{c}{N=15} && \multicolumn{2}{c}{N=28}& N=43\\ 
SFR ($M_{\odot}$/yr) & 868$^{+553}_{-338}$ & 683$^{+560}_{-420}$ && 145$^{+29}_{-33}$ & 74$^{+26}_{-27}$ & 397\\
All sources          & \multicolumn{2}{c}{N=34} && \multicolumn{2}{c}{N=66} & N=100\\ 
SFR ($M_{\odot}$/yr) & 1176$^{+477}_{-339}$ & 1010$^{+706}_{-503}$  &  &&& 497\\
&&&&&\\
\hline
\end{tabular}
\label{tab:SFR_summary}
\end{table}

\clearpage
\renewcommand\thefigure{\thesection.\arabic{figure}}
\appendix

\section{Appendix}

The scan IDs of the HerMES fields relevant to this work are:

\begin{table}[h]
\vspace*{1cm}
\label{scans_IDs}
\begin{tabular}{l l}
 \toprule
$HerMES~ field$ & $Scan~ IDs$ \\
 \midrule
  Lockman$-$North & \makecell[l]{1342186110, 1342222588, 1342222589, 1342222590, 1342222591, \\ 1342222593, 1342222594 1342222595, 1342222596} \\
  \\
  Lockman$-$Swire & \makecell[l]{1342186108, 1342186109, 1342222588, 1342222589, 1342222590, \\ 1342222591, 1342222593, 1342222594, 1342222595, 1342222596}  \\
  \\
  GOODS$-$North & 1342185536\\
  \\
   B\"{o}otes & \makecell[l]{1342187711, 1342187712, 1342187713, 1342188090, 1342188650, \\ 1342188651, 1342188681, 1342188682, 1342189108}\\
\\
   ELAIS$-$N1 & \makecell[l]{1342187646, 1342187647, 1342187648, 1342187649, 1342187650} \\

                     \bottomrule


 \end{tabular}
\end{table}

\newpage
When searching for detections in the large surveys we considered the following catalogues:
\begin{table}[h]
\vspace*{1cm}
\label{scans_IDs}
\scriptsize
\begin{tabular}{l l l}
 \toprule
$Survey~field$ & $Catalogue~name$ & $Web~link$\\
 \midrule
  HerS & hers\_catalogue\_3sig250\_no\_extended.fits & http://www.astro.caltech.edu/hers/Data4\_Product\_Download.html \\
  \\
  H$-$Atlas SDP &  HATLAS\_SDP\_catalogue.fits & http://www.h-atlas.org/public-data/download\\
  \\
  HerMES Lockman$-$North & \makecell{L3-Lockman-North\_xID250\_DR2.fits.gz\\L3-Lockman-North\_xID350\_DR2.fits.gz\\L3-Lockman-North\_xID500\_DR2.fits.gz} & http://hedam.lam.fr/HerMES/index/all\_files\\
  \\
  HerMES Lockman$-$Swire  &   \makecell{L5-Lockman-SWIRE\_SCAT250SXT\_DR2.fits.gz\\L5-Lockman-SWIRE\_SCAT350SXT\_DR2.fits.gz\\L5-Lockman-SWIRE\_SCAT500SXT\_DR2.fits.gz} & http://hedam.lam.fr/HerMES/index/all\_files\\
  \\
  HerMES GOODS$-$North &  \makecell{L2-GOODS-North\_SCAT250SXT\_DR2.fits.gz\\L2-GOODS-North\_SCAT350SXT\_DR2.fits.gz\\L2-GOODS-North\_SCAT500SXT\_DR2.fits.gz} & http://hedam.lam.fr/HerMES/index/all\_files\\
  \\
  HerMES B\"{o}otes & \makecell{L5-Bootes-HerMES\_SCAT250SXT\_DR2.fits.gz\\L5-Bootes-HerMES\_SCAT350SXT\_DR2.fits.gz\\L5-Bootes-HerMES\_SCAT500SXT\_DR2.fits.gz} & http://hedam.lam.fr/HerMES/index/all\_files\\
  \\
  HerMES ELAIS$-$N1 & \makecell{L5-ELAIS-N1-HerMES\_SCAT250\_DR2.fits.gz\\L5-ELAIS-N1-HerMES\_SCAT350\_DR2.fits.gz\\L5-ELAIS-N1-HerMES\_SCAT500\_DR2.fits.gz} & http://hedam.lam.fr/HerMES/index/all\_files\\

                     \bottomrule


 \end{tabular}
\end{table}

\begin{figure}[t]
\includegraphics[height=18cm]{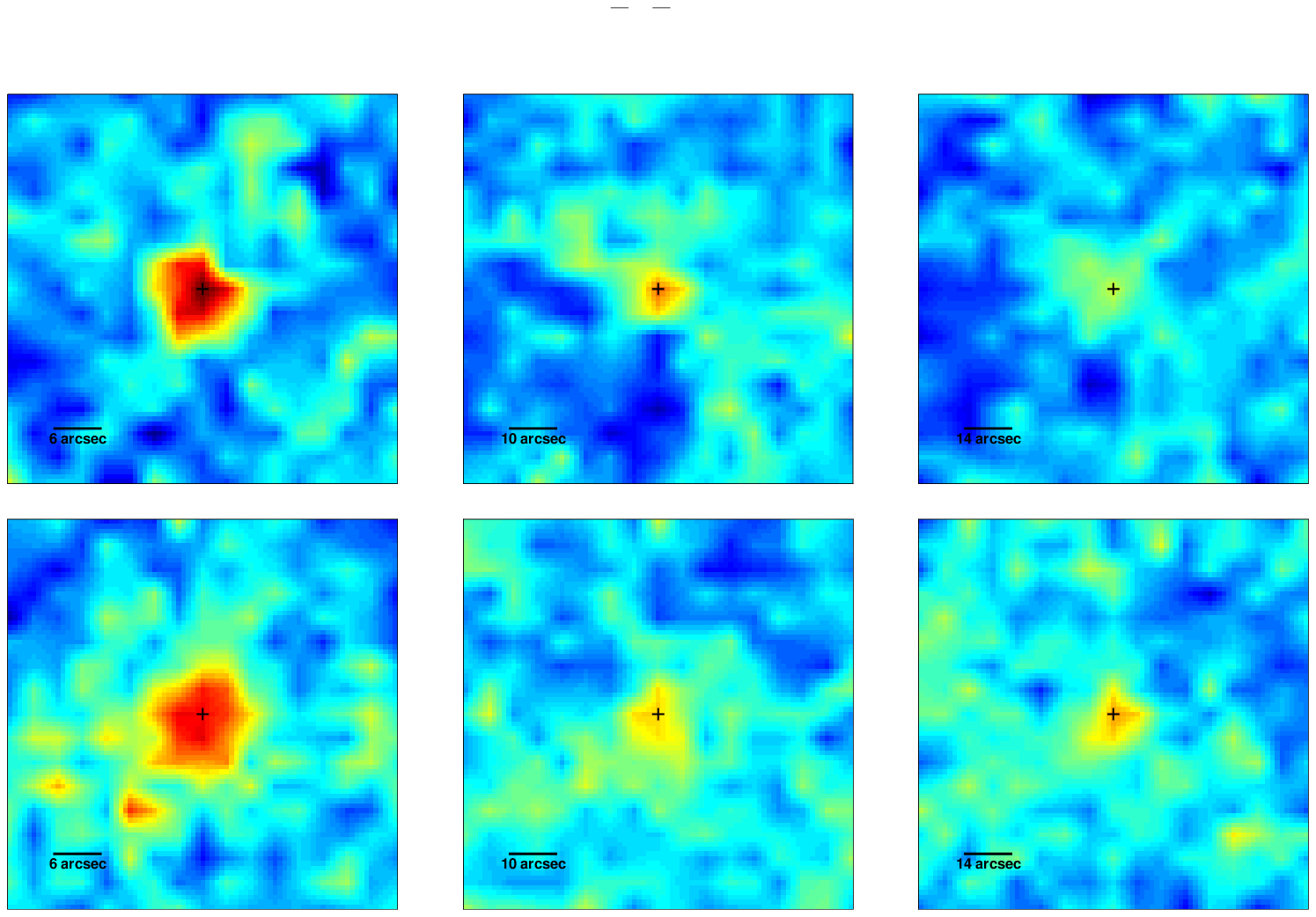}
\caption{Median (top) and mean stacks. In each part, the high luminosity group in shown in the upper row and
the low luminosity group in the lower row.
From left to right: 250 \mic, 350 \mic, and 500 \mic.
The crosses mark the centers of the stacks.
}
\label{fig:stacks}
\end{figure}

\end{document}